\documentclass[a4paper,12pt]{article}
\usepackage{jheppub}
\usepackage{ulem}
\usepackage{slashed}

\newcommand{\psla}{\mbox{$\not{\! p}$}} 
\renewcommand{\Re}{\operatorname{Re}}
\renewcommand{\Im}{\operatorname{Im}}
\def\tot{\text{tot}}
\def\CMS{\text{\tiny CMS}}
\def\MeV{\text{MeV}}
\def\GeV{\text{GeV}}

\def\sqs{\sqrt{s}}

\def\b{\beta}
\def\G{\Gamma}
\def\g{\gamma}
\def\vb{\bar{v}}
\def\cM{\mathcal{M}}
\def\cO{\mathcal{O}}
\def\sW{\scriptscriptstyle{W}}
\def\sZ{\scriptscriptstyle{Z}}
\def\sH{\scriptscriptstyle{H}}
\def\OS{\scriptscriptstyle{\text{OS}}}
\def\CMS{\scriptscriptstyle{\text{CMS}}}

\title{{ The effects of the widths on the one-loop electroweak corrections to  
  the $pp \to WW$ process}}
  
\author[a]{N. Bekheddouma Abdi}
\author[b]{, R. Bouamrane}
\author[c]{and K. Khelifa-Kerfa}

\affiliation[a]{LMA, Département de Physique, Faculté des Sciences Exactes et Informatique\\
	Université Hassiba Benbouali de Chlef - Chlef 02000, Algeria}
	
\affiliation[b]{LEPM, Faculty of Physics, Université des Sciences et de la Technologie d’Oran\\
Mohamed-Boudiaf USTO-MB, BP 1505, EL M’naouer, Oran, Algeria}

\affiliation[c]{LMA, Département de Physique, Faculté des Sciences et de la Technologie \\
Université Relizane - Relizane 48000, Algéria}

\emailAdd{n.bekheddoumaabdi@univ-chlef.dz}
\emailAdd{rachid.bouamrane@univ-usto.dz}
\emailAdd{kamel.khelifakerfa@univ-relizane.dz}

\bigskip 
\abstract{
In this paper, we study the effects of the widths of unstable particles on the one-loop electroweak corrections for the  $pp \to WW$ process at the TeV scale within the framework of the complex mass scheme. We also investigate, for this same process, the unitarity of the theory at high energies.

}
	

\begin{document}
\maketitle
\flushbottom

\section{Introduction}

After the discovery of the Higgs boson in 2012 at CERN,
the particle responsible, via the Brout-Englert-Higgs mechanism, for generating the masses of the particles of the Standard Model (SM), \,we entered the era of Higgs-precision tests of SM. \, Since most known fundamental particles are unstable, it becomes important to consider the width, $\Gamma$, of these particles when evaluating physical observables \cite{CMS-Bauer-2012, CMS-Denner-2015, CMS-Kuksa-2015}. Since The electroweak sector being the most sensitive to unstable particles, such as the weak gauge bosons, the Higgs boson and the top quark, we study, within the framework of the SM, the effects of the widths of the latter on the $pp \to WW$ process.

Knowing that the inclusion of the finite width is not trivial, since it can invoke, following a mixture of perturbative orders, a break in the gauge invariance during calculations of radiative corrections to one-loop \cite{Berends:1969nt, ARGYRES1995339}. Many approaches have been proposed in the past to incorporate the said width into perturbative computations, such as the fixed-width scheme \cite{ARGYRES1995339}, the narrow-width approximation \cite{Baur-1995}, the pole scheme \cite{Beenakker-97,Stuart-91} and the effective field theory \cite{Beneke-2003,Beneke-2004}. But the complex mass scheme \cite{CMS-Denner-2005,CMS-Denner-2006, DENNER2012504, Frederix:2018nkq} is the only one, by construction, which preserves all of the algebraic relations which satisfy gauge invariance in the resonant and non-resonant regions, where the main idea is based on an analytic continuation of the parameters of the Lagrangian SM, which are related to the masses of unstable particles, in the complex plane. Therefore, renormalizability and unitarity are also preserved. This scheme has been implemented in various high energy software such as \texttt{MadGraph5 aMC@NLO} \cite{Alwall:2014hca, Frederix:2018nkq}, \texttt{OpenLoops 2} \cite{Buccioni:2019sur} and \texttt{Recola} \cite{Actis_2017}. The aim of this paper is to study the effect of the width of the weak gauge bosons, the Higgs and the top quark on the process $pp \to WW$ at the one-loop level within the complex mass scheme. 

The paper is organised as follows. In section \ref{sec:CMS} we summarise the main ideas of the CMS. Then, in section \ref{sec:Effects}, we compute the cross section of the process considered herein, i.e.,  $pp \to WW$. up to one-loop, within both the usual on-shell scheme (which we shall be referring to as ``real scheme") and the complex mass scheme. We analyse the results obtained in the said two schemes and discuss the effects of the width on the unitarity. Finally, we draw our conclusions in section \ref{sec:Conclusion}.

\section{Complex mass scheme in a nutshell}
\label{sec:CMS}

The complex mass scheme is a renormalisable scheme which deals properly with unstable particles in all phase space. It was first used in tree-level calculations with W/Z resonances, then generalised to $\mathcal{O}(\alpha)$ \cite{CMS-Denner-2005, CMS-Denner-2006}. 
In this scheme and at the tree level,  the real masses of the weak gauge bosons $W/Z$ and the Higgs boson are changed to complex quantities, defined as the position of the poles of the corresponding propagators which have complex momenta $k$. To preserve the gauge invariance we have to introduce the complex masses everywhere in the Feynman rules, in such a way that the bare Lagrangian remains invariant, particularly in the weak mixing angle;
\begin{equation}
\hat{\mu}_{\sW}^2 = M_{\sW}^2 - \imath M_{\sW} \Gamma_{\sW},
\quad
\hat{\mu}_{\sZ}^2 = M_{\sZ}^2 - \imath M_{\sZ} \Gamma_{\sZ},
\quad
\hat{\mu}_{\sH}^2 = M_{\sH}^2 - \imath  M_{\sH} \Gamma_{\sH},
\end{equation} 
and 
\begin{equation}
\cos^2 \theta_{\sW} \equiv \hat{c}_{\sW}^2 = 1 - \hat{s}_{\sW}^2 = \frac{\hat{\mu}_{\sW}^2}{\hat{\mu}_{\sZ}^2}\simeq \frac{M_W^2}{M_Z^2}\left[ 1 - \imath \left( \frac{\Gamma_W}{M_W}-\frac{\Gamma_Z}{M_Z}\right)\right].
\end{equation}
The hat over the masses and the mixing angles denotes the fact that they are complex-valued.
Since the gauge invariance is not affected by this analytical continuation of the mass in the complex $k^2$ plane: $m \longrightarrow \hat{\mu} = m- \imath \G/2 \longrightarrow \hat{\mu} (k) = m(k) - \imath \Gamma(k)/2$, thus the Ward identity and that of Slavnov-Taylor are preserved. This signifies that the elements of the S-matrix are independent of the gauge parameters.    

Although the introduction of the complex mass in the resonant propagators is trivial,
\begin{align}
\frac{1}{k^2 - M_B^2} &\longrightarrow \frac{1}{k^2 - \hat{\mu}_B^2}\simeq \frac{1}{k^2 - M_B^2\left(1-\imath \frac{\Gamma_B}{M_B}\right)},
\\ 
\frac{1}{\psla-m_F} &\longrightarrow \frac{1}{\psla -\hat{\mu}_F}=\frac{1}{\psla - m_F\left(1-\imath \frac{\Gamma_F}{2 m_F}\right)},
\end{align}
where the subscripts $B$ and $F$ stand for bosons and fermions, respectively, it induces in other regions (such as the weak mixing angle) when passing to the complex mass scheme spurious terms of order $ \mathcal{O}(\alpha) $ in the tree-level amplitude, thus only affecting loop-level calculations. 

To generalise the CMS to the one-loop level while keeping the bare Lagrangian invariant we split the real masses in the latter for the unstable particles into \
renormalised complex masses and complex counter-terms. The resultant Feynman rules enable to perform perturbative calculations exactly as in the usual renormalisation on-shell scheme. To this end, we add and subtract the same imaginary part of each mass of the unstable particle. One of these imaginary parts is incorporated into the free propagator to define the complex mass of the corresponding unstable particle. The other part is introduced into the vertex counter-term. The first term is thus resummed but the second one is not. This prescription does not affect the gauge invariance but may invoke a violation of unitarity of order $\mathcal{O}(\alpha^2)$ in calculations at order $\mathcal{O}(\alpha)$. This is due to the fact that the modified renormalised Lagrangian is not hermitian \cite{Belanger-2006}. Apart from this problem, the complex mass scheme is coherent and gauge invariant in the next-to-leading order (NLO) calculations. Its implementation in a numerical code at one-loop is feasible, since it suffices to redefine the counter-terms by including imaginary parts for two-point functions.       
The complex masses are not only introduced for the gauge bosons but for all unstable particles relevant to the electroweak sector such as the top quark.

\subsection{Complex renormalisation}
\label{sec:ComplexRenor}

In this section we summarise the procedure of the generalised renormalisation which takes into account the complex masses in the 't Hooft-Feynman gauge in a straight-forward way \cite{CMS-Denner-2005}.

Since the bare Lagrangian is unaffected the complex masses of the gauge bosons, $W$ and $Z$, are introduced in the latter after decomposing the bare real masses into renormalised complex masses and complex counter-terms: 
\begin{equation}
M_{W,0}^2 = \hat{\mu}_{W}^2 + \delta \hat{\mu}_W^2, 
\quad\quad\quad
M_{Z,0}^2 = \hat{\mu}_{Z}^2 + \delta \hat{\mu}_Z^2,
\end{equation}
where we note that the following consistency condition should be respected:
\begin{equation}
\Im(\hat{\mu}_{V}^2) = - \Im(\delta \hat{\mu}_{V}^2). 
\end{equation}
In the above equation the subscript $0$ labels the bare quantities, and $V$ stands for $W$ or $Z$ bosons. In a similar fashion, we observe that the renormalised gauge fields are related to the bare ones via the following relations:
\begin{align}
W_0^\pm = \left(1 + \frac{1}{2} \delta \mathcal{Z}_W \right) W^\pm,
\end{align}
\begin{eqnarray}
\left( 
\begin{array}{c}
Z_0 \\
A_0
\end{array}
\right)
=
\left( 
\begin{array}{cc}
1 + \frac{1}{2} \delta \mathcal{Z}_{ZZ} & \frac{1}{2} \delta \mathcal{Z}_{ZA} \\
\frac{1}{2} \delta \mathcal{Z}_{AZ} & 1 + \frac{1}{2} \delta \mathcal{Z}_{AA}
\end{array}
\right)
\left( 
\begin{array}{c}
Z\\
A
\end{array}
\right).
\end{eqnarray}

Since the renormalisation conditions are the same for stable and unstable particles, that is, the position of the poles of the propagator equals the square of the physical mass and the residue of the propagator equals 1, they assume similar forms  in CMS as those in the usual scheme but without taking the real part of the renormalised transverse self energy (T):
\begin{eqnarray}\label{ren-conds}
\bar{\Sigma}_T^W (\hat{\mu}_W^2) = 0, &\quad\quad &\bar{\Sigma}_T^{ZZ} (\hat{\mu}_Z^2) = 0, \nonumber \\
\bar{\Sigma}_T^{AZ} (0) = 0, &\quad\quad &\bar{\Sigma}_T^{AZ} (\hat{\mu}_Z^2) = 0, \nonumber \\
\bar{\Sigma}_T^{\prime W} (\hat{\mu}_W^2) = 0, &\quad\quad &\bar{\Sigma}_T^{\prime ZZ} (\hat{\mu}_Z^2) = 0, \quad\quad\bar{\Sigma}_T^{\prime AA} (0) = 0, 
\end{eqnarray}
where the prime means differentiation with respect to the argument and the bar on the $\Sigma$'s indicates that they are renormalised. The first two terms in Eq. \eqref{ren-conds} fix the counter-terms of the masses of the $W$ and $Z$ bosons, while the last five terms fix the counter-terms of their fields. Knowing that the generalised renormalised transverse self energies are the same as the usual on-shell scheme with the replacement of the real masses and counter-terms by their complex counterparts. The solutions of the said conditions \eqref{ren-conds} are as follows:
\begin{align}\label{solutions-RC}
& \delta\hat{\mu}_W^2 = \Sigma_T^W(\hat{\mu}_W^2), \quad\quad \delta\mathcal{Z}_W = - \Sigma_T^{\prime W}(\hat{\mu}_W^2), 
\notag\\
& \delta \mathcal{Z}_{ZA} = \frac{2}{\hat{\mu}_Z^2} \Sigma_T^{AZ}(0), \quad 
\delta \mathcal{Z}_{AZ} = - \frac{2}{\hat{\mu}_Z^2} \Sigma_T^{AZ}(\hat{\mu}_Z^2), \qquad 
\notag \\
& \delta \mathcal{Z}_{W} = - \Sigma_T^{'W}(\hat{\mu}_W^2), \quad 
\delta \mathcal{Z}_{ZZ} = - \Sigma_T^{'ZZ}(\hat{\mu}_Z^2), \quad
\delta \mathcal{Z}_{AA} = - \Sigma_T^{'AA}(0).
\end{align}  
It requires analytical continuation to compute the above two-point functions with complex arguments. To avoid this complication we expand the self energies around real arguments. To see how to transform the renormalised self energies and the solutions of the renormalisation conditions, we concentrate on the case of the W gauge boson. We have 
\begin{equation}\label{eq:WRenSelfEnergy-Expand}
\bar{\Sigma}_T^W (k^2) = \Sigma_T^W (k^2) - \delta \hat{\mu}_W^2 + (k^2 - \hat{\mu}_W^2) \delta\mathcal{Z}_W,
\end{equation}
with 
\begin{align}\label{eq:Expansion}
\Sigma_T^W(\hat{\mu}_W^2) = \Sigma_T^W(M_W^2) + \left(\hat{\mu}_W^2 - M_W^2 \right) \Sigma^{'W}_T(M_W^2) + \mathcal{O}(\alpha^3). 
\end{align}
Neglecting the terms at $\mathcal{O}(\alpha^3)$ and beyond we obtain the modified solutions, which when inserted into \eqref{eq:WRenSelfEnergy-Expand} results in a form of the renormalised transverse self energy that resembles that of the usual on-shell scheme but without taking the real part of the solutions: 
\begin{align}\label{solutions-RC}
\bar{\Sigma}_T^W (k^2) = \Sigma_T^W (k^2) - \delta M_W^2 + (k^2 - M_W^2) \delta\mathcal{Z}_W
\notag \\
\delta M_W^2 = \Sigma_T^W(M_W^2), \quad\quad \delta\mathcal{Z}_W = - \Sigma_T^{\prime W}(M_W^2).
\end{align}
While in the on-shell scheme the self-energies are calculated with  real renormalised masses, in the CMS self-energies, Eq. \eqref{solutions-RC}, ought to be calculated with complex masses, but with real squared momenta. This enables us to avoid analytic continuation in the momentum space.

In order to correctly address resonances at order $\mathcal{O}(\alpha)$ one ought to take into account the $W$ boson width, $\Gamma_W$, including $\mathcal{O}(\alpha)$ corrections. This may be obtained in an iterative way from the following equation: 
\begin{equation}
M_W\Gamma_W = \Im \left\{\Sigma_T^W(M_W^2) \right\} - M_W\Gamma_W \Re \left\{\Sigma_T^{\prime W}(M_W^2)\right\} + \mathcal{O}(\alpha^3).
\end{equation}
This latter equation can be easily deduced from the imaginary part of \eqref{eq:Expansion}. 
Furthermore, the complex weak mixing angle is renormalised as follows: 
\begin{equation}
\frac{\delta \hat{c}_{\sW}}{\hat{c}_{\sW}} 
= \frac{1}{2} \left(  \frac{\delta \hat{\mu}_W^2}{\hat{\mu}_W^2} - \frac{\delta \hat{\mu}_Z^2}{\hat{\mu}_Z^2} \right)  = \frac{1}{2} \left[ \frac{ \Sigma_T^W(\hat{\mu}_W^2) }{\hat{\mu}_W^2} - \frac{ \Sigma_T^Z (\hat{\mu}_Z^2) }{\hat{\mu}_Z^2}  \right].
\end{equation}

For the Higgs boson, the renormalisation constants can be approached in the same manner as before;
\begin{equation}
M_{H,0}^2 = \hat{\mu}_{H}^2 + \delta \hat{\mu}_H^2, \qquad\qquad\qquad\qquad\qquad\qquad\quad
\end{equation}
with
\begin{align}
\delta \hat{\mu}_H^2 &= \Sigma^H(\hat{\mu}_H^2),  \notag\\
&= \Sigma^H(M_H^2)+ (\hat{\mu}_H^2-M_H^2)\Sigma^{\prime H} (M_H^2) + \mathcal{O}(\alpha^3). 
\notag \\
\delta \mathcal{Z}_H &= - \Sigma^{'H}(\hat{\mu}_H^2), \notag \\
&= -\Sigma^{\prime H}(M_H^2)+ \mathcal{O}(\alpha^2). 
\end{align}
Hence the renormalised self energy for the Higgs boson up to the $\mathcal{O}(\alpha^2)$  can be written as:
\begin{equation}
\bar{\Sigma}^H (k^2) = \Sigma^H(k^2) - \delta M_H^2 + (k^2 - M_H^2)\, \delta \mathcal{Z}_H  \qquad\qquad
\end{equation}
where
\begin{equation}
\delta M_H^2 = \Sigma^H(M_H^2), \qquad \delta \mathcal{Z}_H = - \Sigma^{\prime H}(M_H^2), \qquad\quad
\end{equation}

Since in the CMS, the complex masses are not introduced solely for gauge bosons and the Higgs boson but for all unstable particles such as the top quark. The renormalisation of this latter may be treated in a similar manner as before with the introduction of its complex mass and counter-term:
\begin{align}
 \hat{\mu}_t^2 &= m_t^2 - \imath m_t \, \Gamma_t, \quad\quad \notag \\
 m_{t,0} &= \hat{\mu}_t + \delta \hat{\mu}_t.  
\end{align}    
The generalised renormalisation constants are determined by:
\begin{align}\label{eq:TopSelfEnergy}
\delta \hat{\mu}_t &= \frac{\hat{\mu}_t}{2} \left[ \Sigma^{t,R}(\hat{\mu}_t^2) + \Sigma^{t,L}(\hat{\mu}_t^2) + 2 \Sigma^{t,s}(\hat{\mu}_t^2)  \right], 
\notag\\
 \delta \mathcal{Z}_{t, \sigma} &= - \Sigma^{t,\sigma}(\hat{\mu}_t^2)  - \hat{\mu}_t^2 \left[ \Sigma^{'t,R}(\hat{\mu}_t^2) + \Sigma^{'t,L}(\hat{\mu}_t^2) + 2 \Sigma^{'t,s}(\hat{\mu}_t^2)  \right],
\end{align}    
where $\sigma = R, L$ indicates the left- and right-handed components of the top self-energy, $\Sigma^t(p)$, following the convention of Ref.  \cite{denner1993techniques}. The generalised renormalised self-energy of the top quark is determined by: 
\begin{multline}
 \bar{\Sigma}^t(p) = \left(\Sigma^{t,R}(p^2)  + \delta \mathcal{Z}_{t,R}\right) \slashed{p}\, P_R + \left(\Sigma^{t,L}(p^2)  + \delta \mathcal{Z}_{t,L}\right) \slashed{p}\, P_L + \\
+\hat{\mu}_t \left[\Sigma^{t,s} - \frac{1}{2}  \left( \delta \mathcal{Z}_{t,R} + \delta \mathcal{Z}_{t,L} \right) - \frac{\delta \hat{\mu}_t}{\hat{\mu}_t} \right], 
\end{multline} 
where the factors $P_{R,L}$ are defined below. 
This becomes, after the expansion of the self-energies of \eqref{eq:TopSelfEnergy} around the real mass $m_t^2$ and negligence of higher order terms:
\begin{multline}
 \bar{\Sigma}^t(p) = \left(\Sigma^{t,R}(p^2)  + \delta \mathcal{Z}_{t,R}\right) \slashed{p}\, P_R + \left(\Sigma^{t,L}(p^2)  + \delta \mathcal{Z}_{t,L}\right) \slashed{p}\, P_L + \\
+\hat{\mu}_t \left[\Sigma^{t,s} - \frac{1}{2}  \left( \delta \mathcal{Z}_{t,R} + \delta \mathcal{Z}_{t,L} \right) - \frac{\delta m_t}{m_t} \right],
\end{multline} 
with 
\begin{align}\label{eq:TopSelfEnergy-2}
\delta m_t &= \frac{m_t}{2} \left[ \Sigma^{t,R}(m_t^2) + \Sigma^{t,L}(m_t^2) + 2 \Sigma^{t,s}(m_t^2)  \right], 
\notag\\
 \delta \mathcal{Z}_{t, \sigma} &= - \Sigma^{t,\sigma}(m_t^2)  - m_t^2 \left[ \Sigma^{'t,R}(m_t^2) + \Sigma^{'t,L}(m_t^2) + 2 \Sigma^{'t,s}(m_t^2)  \right].
\end{align}   

Before ending this section it is important to recall that the masses of the external particles for a given process must be real, i.e., they are considered as stable particles as had been shown by Veltman \cite{VELTMAN1963186}. Moreover, the same particles should not be taken in the same process as internally unstable (resonances) and externally stable, because they cannot be treated simultaneously by two different schemes (usual on-shell and CMS) \cite{Denner:2019vbn}.

\section{Width effects on $p p \to W W$ at one loop}
\label{sec:Effects}

\subsection{Cross-section at lowest order}
\label{sec:Born}

\begin{figure}[h!]
\centering
\includegraphics[scale=0.45]{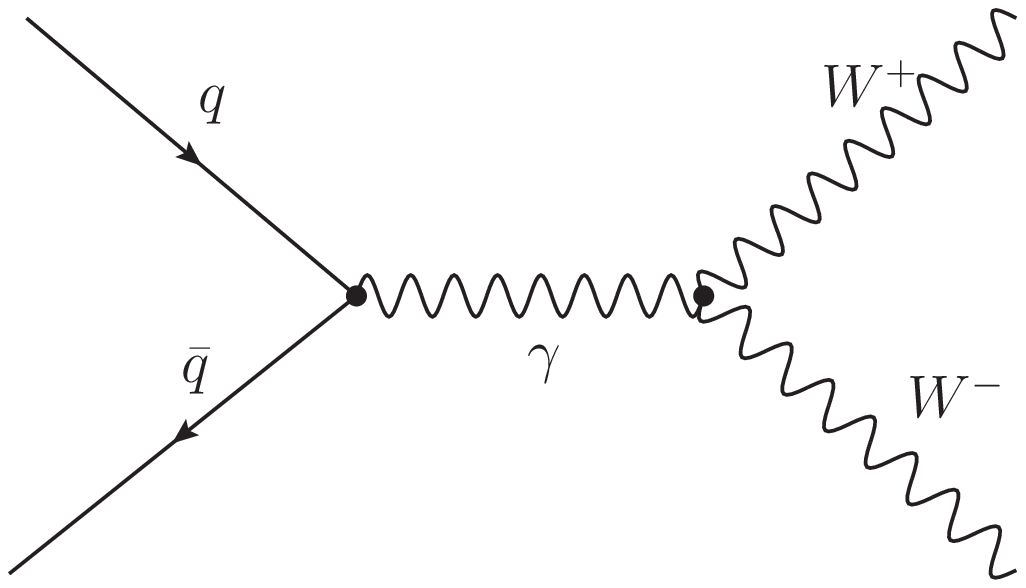}
\includegraphics[scale=0.45]{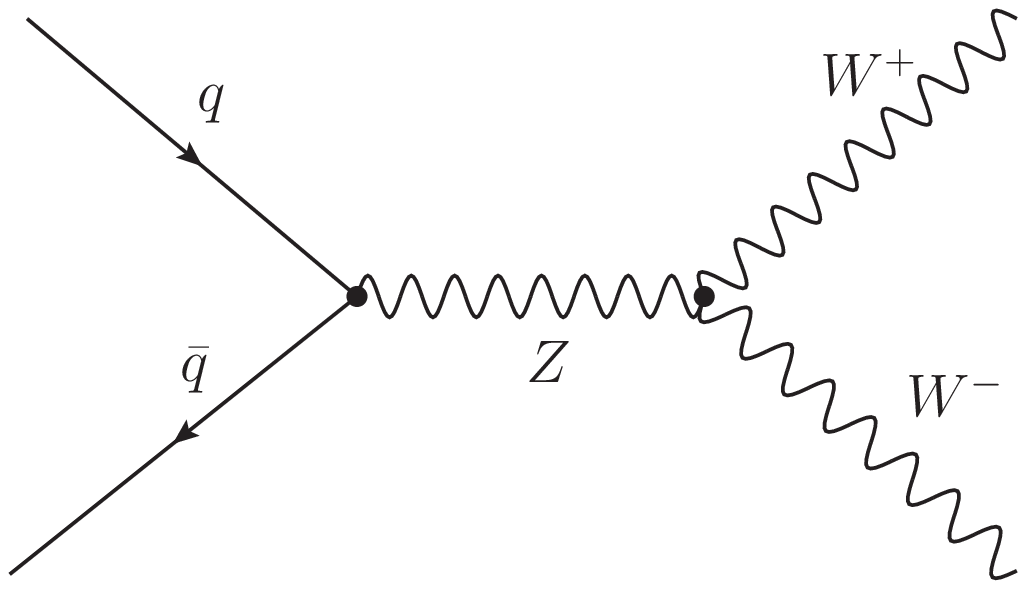}
\includegraphics[scale=0.45]{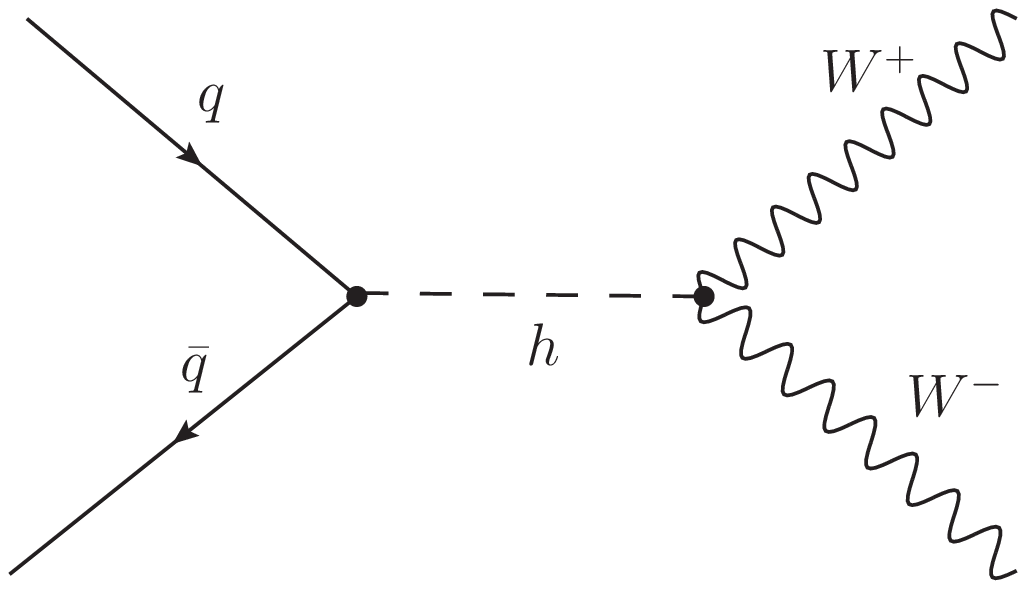}
\includegraphics[scale=0.45]{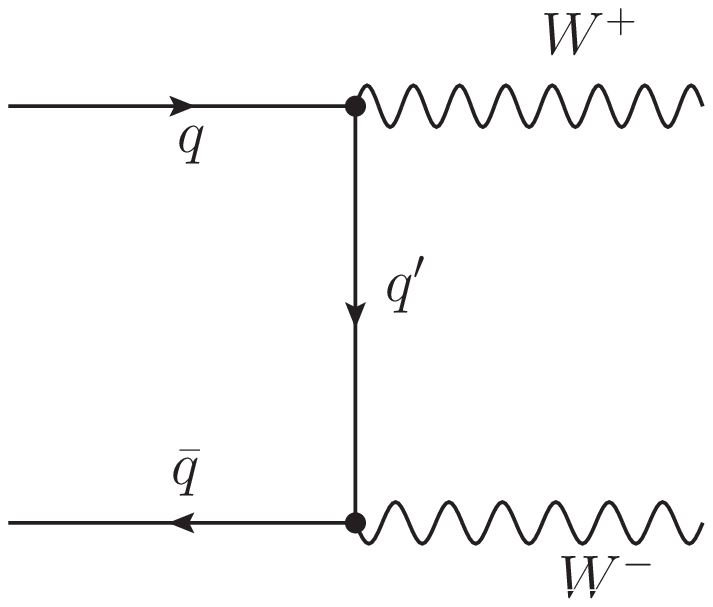}
\caption{Feynman diagrams for the Born process $q \bar{q} \to W^+ W^-$.} \label{fig:Feyn-diags}
\end{figure}

We shall be investigating the process 
\begin{align}
 P(p_1) + P(p_2) \to W^+(k_1, \lambda_1) + W^-(k_2, \lambda_2),
\end{align}
where $P_i, k_i$ are the momenta of the protons and the W bosons respectively, and $\lambda_i$ are the polarisations ($\lambda_i = 0$ for longitudinal polarisations, referred to as $L$, $\lambda_i = \pm 1$ for transverse polarisations, referred to as $T$, and non-polarised referred to by $U$). The tree-level Born Feynman diagrams corresponding to our process are shown  in Fig. \ref{fig:Feyn-diags}.

The four-momenta of the protons and bosons are given by:
\begin{align}
 p_1 = \frac{\sqs}{2}(1,0,0,\b_q), &\qquad  k_1 = \frac{\sqs}{2}(1,\b \sin\theta,0,\b \cos\theta), \notag\\ 
 p_2 = \frac{\sqs}{2}(1,0,0,-\b_q), &\qquad  k_2 = \frac{\sqs}{2}(1,-\b \sin\theta,0,-\b \cos\theta),  
\end{align}  
where $\b = \sqrt{1- 4M_W^2/s}$, $\b_q = \sqrt{1 - 4 m_q^2/s}$, $M_W$ is the $W$ boson's mass, $m_q$ is the quark's mass and $\theta$ is the scattering angle in the centre-of-mass of the system, with $p_1^2 = p_2^2= m_q^2$ and $k_1^2 = k_2^2 = M_W^2$. The {\it longitudinal} and {\it transverse} polarisation vectors of the final bosons read:
\begin{align}\label{eq:PolVectors}
 \epsilon^1_L(0) = \frac{\sqs}{2 M_W}(\b, \sin\theta, 0, \cos\theta), &\qquad
 \epsilon^1_T(\pm) = (0, \cos\theta, \mp \imath, -\sin\theta)/\sqrt{2}. 
  \notag\\
 \epsilon^2_L(0) = \frac{\sqs}{2 M_W}(\b, -\sin\theta, 0, -\cos\theta). &\qquad
   \epsilon^2_T(\pm) = (0,-\cos\theta, \mp \imath, \sin\theta)/\sqrt{2}.
\end{align}
It is worth mentioning that only the longitudinal polarisation is massive. In CMS, the widths of the unstable particles are introduced, at tree-level, through propagators of $Z, W, H, t$ and the weak mixing angle. We shall only be considering, in this section, the mode $p p \to W^+_L W^-_L$ to analytically study the effect of the widths on the tree-level amplitude in the CMS framework, since this mode is affected by the width $\G_W$ unlike the transverse mode. We then compute, using \texttt{MadGraph5 aMC@NLO}, the Born cross-sections for different polarisations of the $W$ gauge boson. 

To see that the unitarity is preserved in the case where the masses of the unstable particles are taken real at high energies, that is in the usual {\it on-shell} (OS) scheme, we introduce in the calculations of the amplitudes of the Feynman diagrams (Fig. \ref{fig:Feyn-diags}) the variable $x = s/4 M_W^2$ with $x \gg 1$. We find, with the aid of \texttt{FeynCalc} \cite{Romao-2016}, the following expression for the total amplitude at high energy:
\begin{align}\label{eq:Amps_sum}
 \cM^{\OS}_{\text{tot}} &= \cM_\g^s + \cM_Z^s + \cM_H^s + \cM_q^t
 \notag\\
 &=
\frac{e^2 (1+2 T_3^f) M_Z^2}{4 \left(M_Z^2 - M_W^2\right) M_W^2} \, \vb(p_2) \left[ m_q(P_R-P_L) - 2 \slashed{q}_1 P_L \right] u(p_1)  + \cO(1/x),
\end{align}
where $M_W = s_W \, M_Z$, $P_{R,L} = (1 \pm \gamma_5)/2$ and $T_3^f = +1/2 (-1/2)$ for $u,c,t (d,s,b)$ quarks.

To study the effect of the decay width of the unstable massive particles ($W, Z, H $ and $t$) of figure \ref{fig:Feyn-diags} on the unitarity of the amplitudes at tree-level and analyse the obtained results at high energies with respect to real mass scheme (OS), we implement the CMS in the process of longitudinal $W$ boson pair production $ p p \to W_L W_L$ (since its polarisation vectors depend on the mass). The expressions of the various amplitudes of the Feynman diagrams \ref{fig:Feyn-diags} following the prescriptions of Denner \cite{Denner:2019vbn} are reported in appendix \ref{app:Amps_CMS} (Eqs. \eqref{eq:Amps_HEL_CMS1}--\eqref{eq:Amps_HEL_CMS4}). The resultant total amplitude is as follows:      
\begin{subequations}\label{eq:Amps_sum_CMS}
\begin{align}
 \cM_{\tot}^{\CMS}  &=  \Re\{ \cM_{\tot}^{\CMS} \} + \imath \Im \{ \cM_{\tot}^{\CMS} \},
\end{align}
where 
\begin{align}
 \Re\{ \cM_{\tot}^{\CMS} \}  &= \frac{e^2 (1+2 T_3^f) M_Z^2 }{4 (M_Z^2 - M_W^2) M_W^2} \, \vb(p_2) \Big[ m_q(P_R-P_L) - 2 \slashed{q}_1 P_L \Big] u(p_1) + \cO(1/x),
 \notag\\
 \Im \{ \cM_{\tot}^{\CMS} \}  &= \frac{e^2 (1+2 T_3^f) \, \G_Z M_Z }{4 (M_Z^2 - M_W^2)^2} \, \vb(p_2) \left[ m_q(P_R-P_L) - 2 \slashed{q}_1 P_L \right] u(p_1) + \cO(1/x).
\end{align}
\end{subequations}

The following points are to be noticed: 
\begin{itemize}
 \item If we set all widths of all internal particles introduced in our process, $\G_W = \G_Z = \G_H = \G_t = 0$, then the real parts of the amplitudes \eqref{eq:Amps_HEL_CMS1}--\eqref{eq:Amps_HEL_CMS4} reduce to those of the real scheme (OS), while the imaginary parts vanish. 
 
\item The real part of the total amplitude $\cM_{\tot}^{\CMS}$ is not affected by the widths of the internal particles, $Z, H$ and $t$, which cancel out in the final expression. The said real part equals the total amplitude in the real scheme; $  \Re\{ \cM_{\tot}^{\CMS} \} =  \cM_{\tot}^{\OS}$, whilst, the imaginary part of  $\cM_{\tot}^{\CMS}$ is affected by the width $\G_Z$ of the internal boson $Z$. It is proportional to $ \cM_{\tot}^{\OS}$: 
\begin{align}
  \Im \{ \cM_{\tot}^{\CMS} \} = \frac{M_W^2\, \G_Z \, M_Z}{M_Z^2 - M_W^2} \, \cM_{\tot}^{\OS}. 
\end{align}
 
 \item The total amplitude depends on the mass and width of the internal boson $Z$ and the masses of the external particles $M_W$ and $m_q$. The other masses, $M_H$ and $m_t$, and their respective widths, $\G_H$ and $\G_t$, cancel out.
 
\item The effect of the width $\G_Z$ of the internal gauge boson $Z$ on the amplitude at tree-level at high energy is around $2\%$. In effect, the ratio of the amplitudes is defined by 
\begin{align}
 \frac{\delta \cM_{\tot}}{\cM_{\tot}^{\OS}} = \frac{\cM_{\tot}^{\OS} - \cM_{\tot}^{\CMS}}{\cM_{\tot}^{\OS}} = -\imath \, \frac{M_W^2}{M_Z^2 - M_W^2} \, \frac{\G_Z}{M_Z}.  
\end{align}   
 
 \item Following the instructions of Denner \cite{Denner:2019vbn} we have taken, in all amplitudes of Feynman diagrams, the width of the external gauge boson $\G_W = 0$. We may, however, see the effect of $\G_W$ on the tree-level amplitude at high energy by introducing it at the level of longitudinal polarisation vectors and the weak mixing angle. We find, as before, that only the imaginary part of  $\cM_{\tot}^{\CMS}$ is affected by the widths $\G_Z$ and $\G_W$ (Eqs. \eqref{eq:Amps_HEL_CMS5} -- \eqref{eq:Amps_HEL_CMS8}). It is not affected, however, by the widths of the Higgs boson and the top quark. 
\begin{align}\label{eq:GW-Effect}
 \Re\{ \cM_{\tot}^{\CMS} \} &= \cM_{\tot}^{\OS}, 
\notag\\
 \Im\{ \cM_{\tot}^{\CMS} \}  &= \frac{e^2 (1+2T_3^f) \, M_Z}{2(M_Z^2-M_W^2)^2} \,\, \vb(p_2)\, \Biggl\{ \, \frac{\G_Z}{M_Z} 
\, \Biggl[ m_q \, \left( 1-2 P_L \right) - 2 \slashed{q_1} P_L  \Biggr] +
 \notag\\
 &  + \frac{\G_W}{M_W} \, \frac{1}{M_Z}  \Bigg[ m_q \left(\frac{T_3^f}{1+2 T_3^f} - P_L \right)- \slashed{q_1}\Bigg( 2 \frac{Q_f}{M_Z} \left(M_Z^2-M_W^2\right)  +
 \notag\\
 & + P_L \Bigg) \Bigg]\Biggr\} \times u (p_1) + \cO(1/x).
 \end{align}
The above equations \eqref{eq:GW-Effect} reduce to those in Eq. \eqref{eq:Amps_sum_CMS} if $\G_W = 0$. 
\item The total tree-level amplitude at high energy is finite in the case where the external bosons are considered either as stable or unstable, thus respecting the unitarity condition.  
\end{itemize}

\subsection{Numerical results}

To perform the numerical calculations of the cross sections at tree-level both in CMS and OS schemes, we use the fixed-order Monte Carlo program  \texttt{MadGraph5\_aMC@NLO}. The following input parameters have been implemented:
\begin{align}\label{eq:Parameters}
M_H  &= 125\, \GeV, \qquad  M_Z = 91.188\, \GeV, \qquad M_W = 80.401\, \GeV, \notag\\
\Gamma_H &= 0.008 \,\GeV, \quad\,\,\, \Gamma_Z= 2.4952\, \GeV, \qquad \,\,\Gamma_W= 2.092698\, \GeV, \notag\\
\alpha^{-1} &= 137.0359895, \,\,\,\, m_t = 173.2  \GeV , \qquad \quad\,\,\, \Gamma_t = 1.3 \GeV .
\end{align}

\begin{figure}[ht]
	\centering
	\includegraphics[scale=0.59]{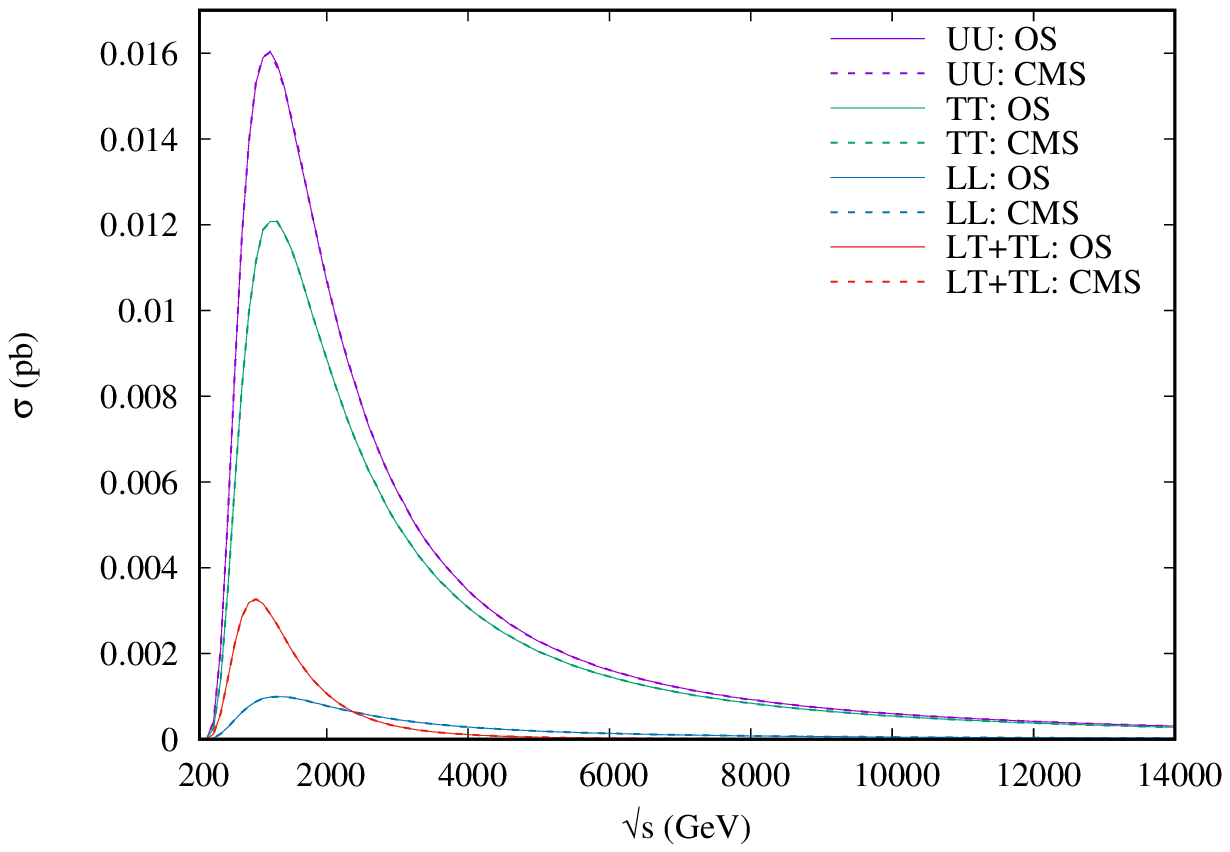}
	\includegraphics[scale=0.59]{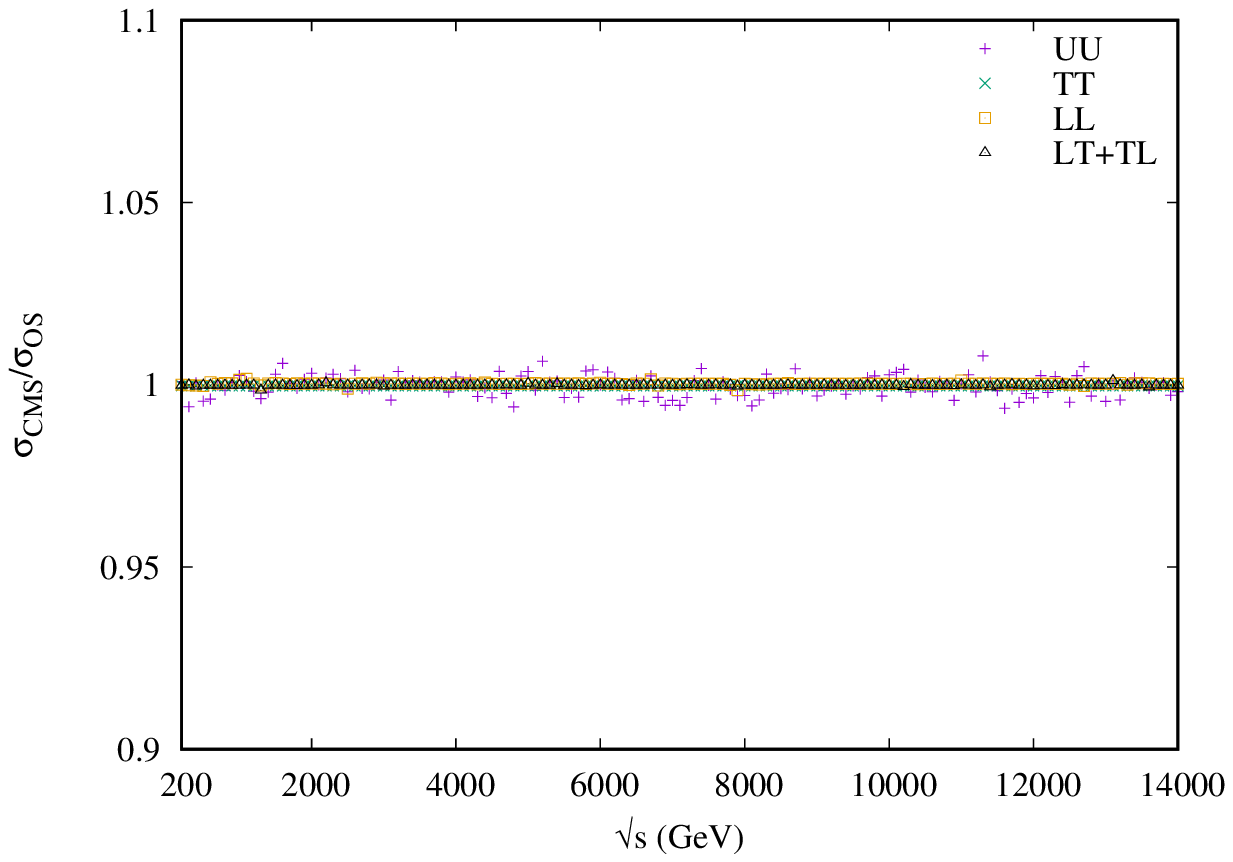}
	\caption{(Left) plots for real (OS) and CMS cross sections as a function of the scattering energy $\sqrt{s}$ for various polarisations of the $W^+ W^-$ bosons. (Right) ratio of the real (OS) and CMS cross sections.  } \label{fig:Born-all}
\end{figure}

In Figure \ref{fig:Born-all}, we present, for different polarisations of the external gauge bosons, the effect of the widths $\G_Z, \G_W, \G_H$ and $\G_t$ on the Born cross sections in the CMS framework in comparison with OS.  
We have added here the effect of the width of the $W$ boson since it does not appear at the internal of Feynman diagrams at tree-level (Fig.  \ref{fig:Feyn-diags}). Hence it does not pose the problem highlighted by Denner \cite{Denner:2019vbn} at one-loop, which states that a given particle cannot be treated by two different schemes when it is taken to be stable externally  and unstable internally.   

For different combinations of polarisations of external bosons, LL, TT and LT + TL, the cross sections behave as $1/s$ at high energy in the real scheme and preserves such behaviour in CMS. These results confirm that the instability of internal and external particles does not affect unitarity at tree-level. We note the following points for the cross sections in both schemes: they are of the same order for all energies used; the maximum is around $1000\,$GeV; in the range $4000 - 6000\,$GeV, the contributions of the LL and LT+TL modes are almost null, whilst the TT and UU modes have non-vanishing values. The TT mode is dominant and makes the principal contribution for the UU mode.      

\begin{figure}[h!]
	\centering
	\includegraphics[scale=0.59]{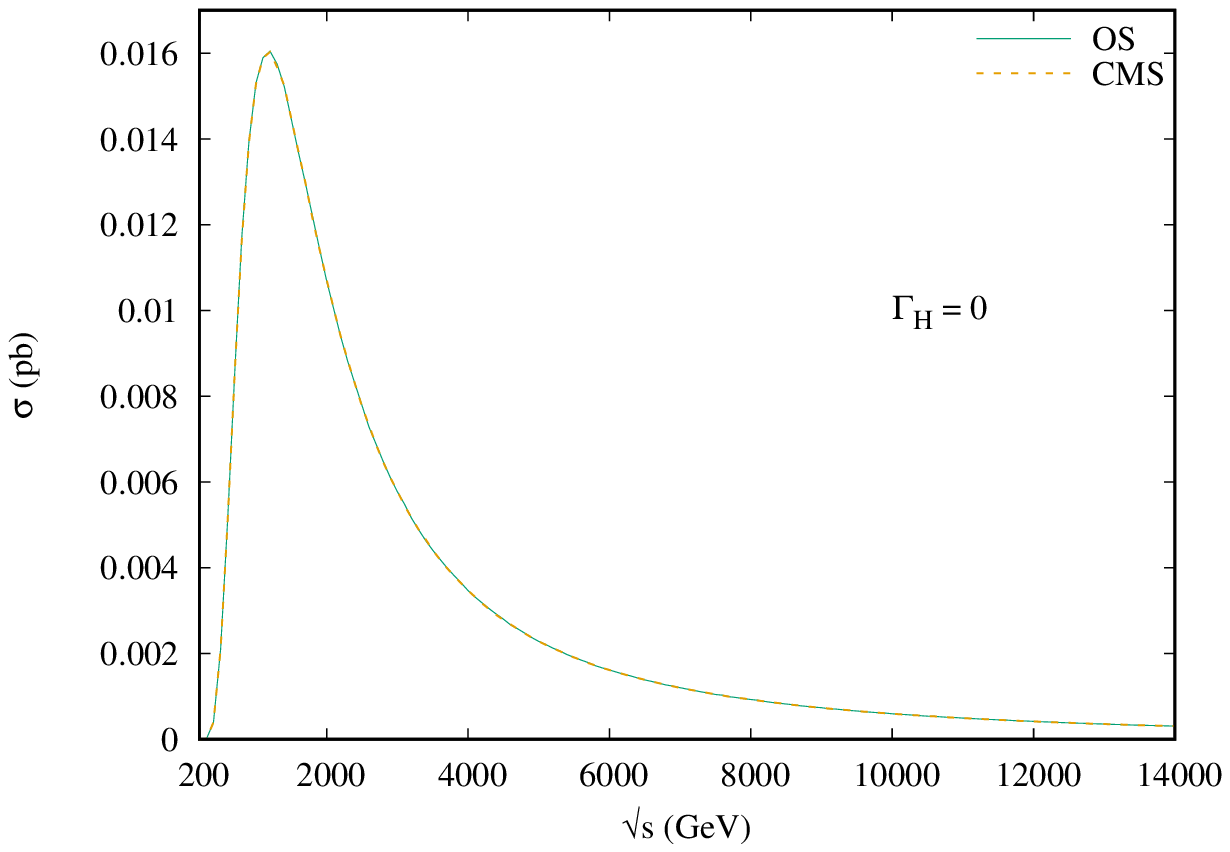}
	\includegraphics[scale=0.59]{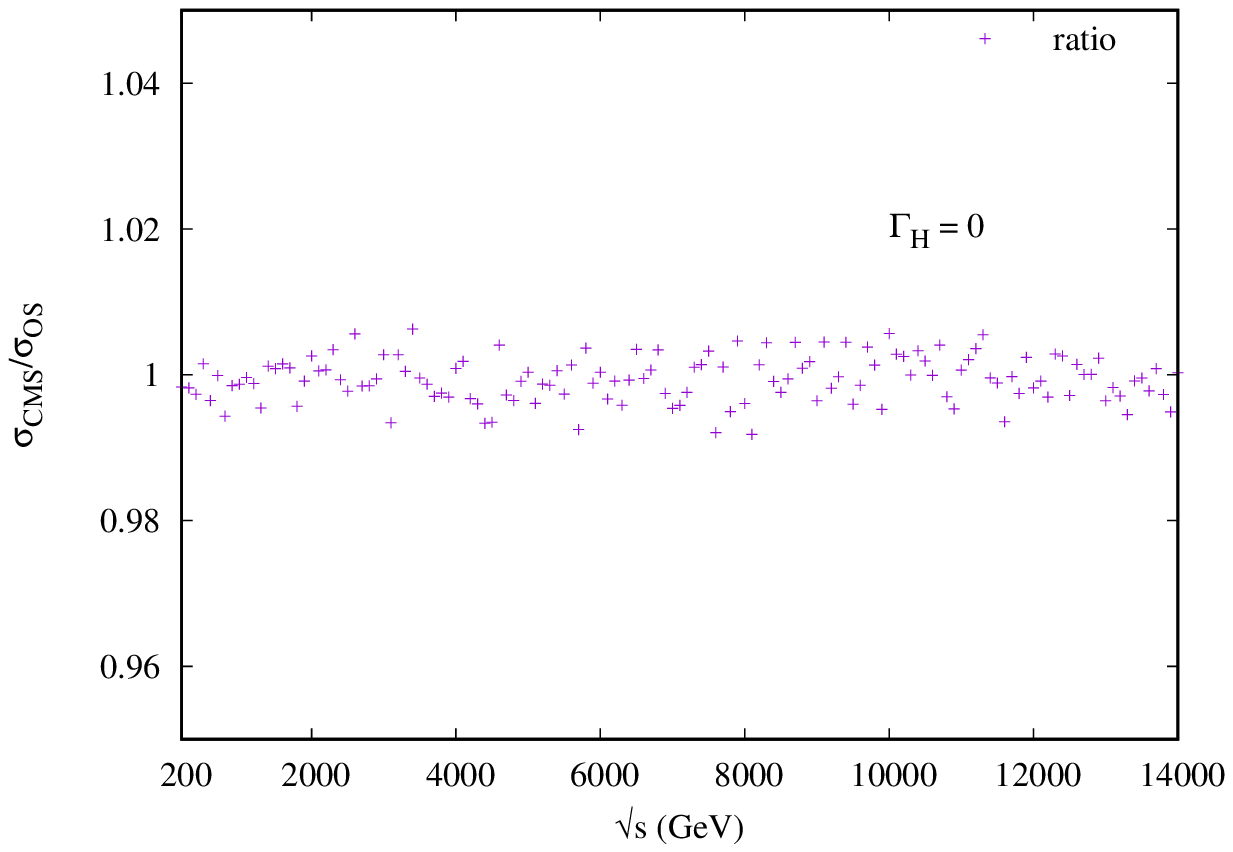}
	\caption{Cross sections in OS and CMS at tree-level for $p p \to W_U W_U$ as a function of $\sqrt{s}$ for the case $\G_H = 0$ (left) and their ratio (right).} \label{fig:Born-GH0}
\end{figure}

\begin{figure}[h!]
	\centering
	\includegraphics[scale=0.59]{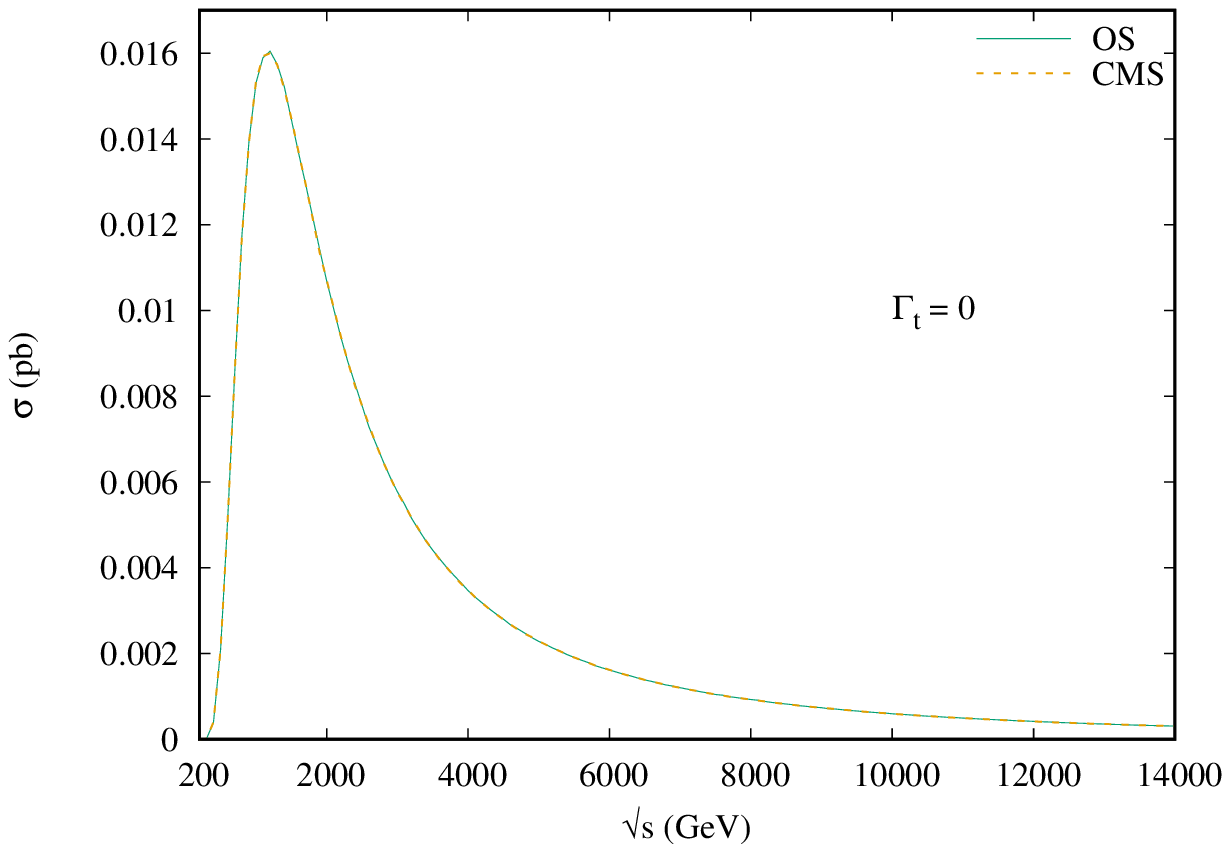}
	\includegraphics[scale=0.59]{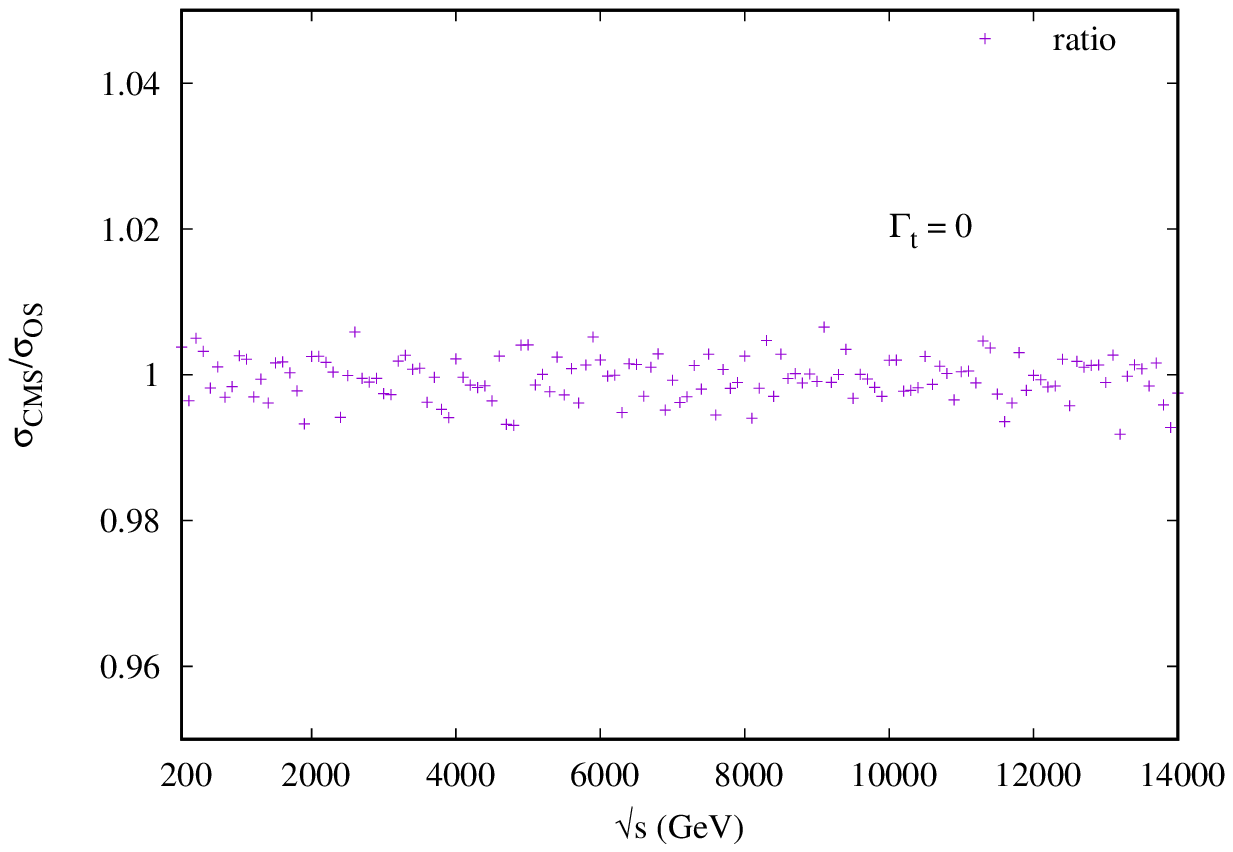}
\caption{Cross sections in OS and CMS at tree-level for $p p \to W_U W_U$ as a function of $\sqrt{s}$ for the case $\G_t = 0$ (left) and their ratio (right).}\label{fig:Born-GT0}
\end{figure}

\begin{figure}[h!]
	\centering
	\includegraphics[scale=0.59]{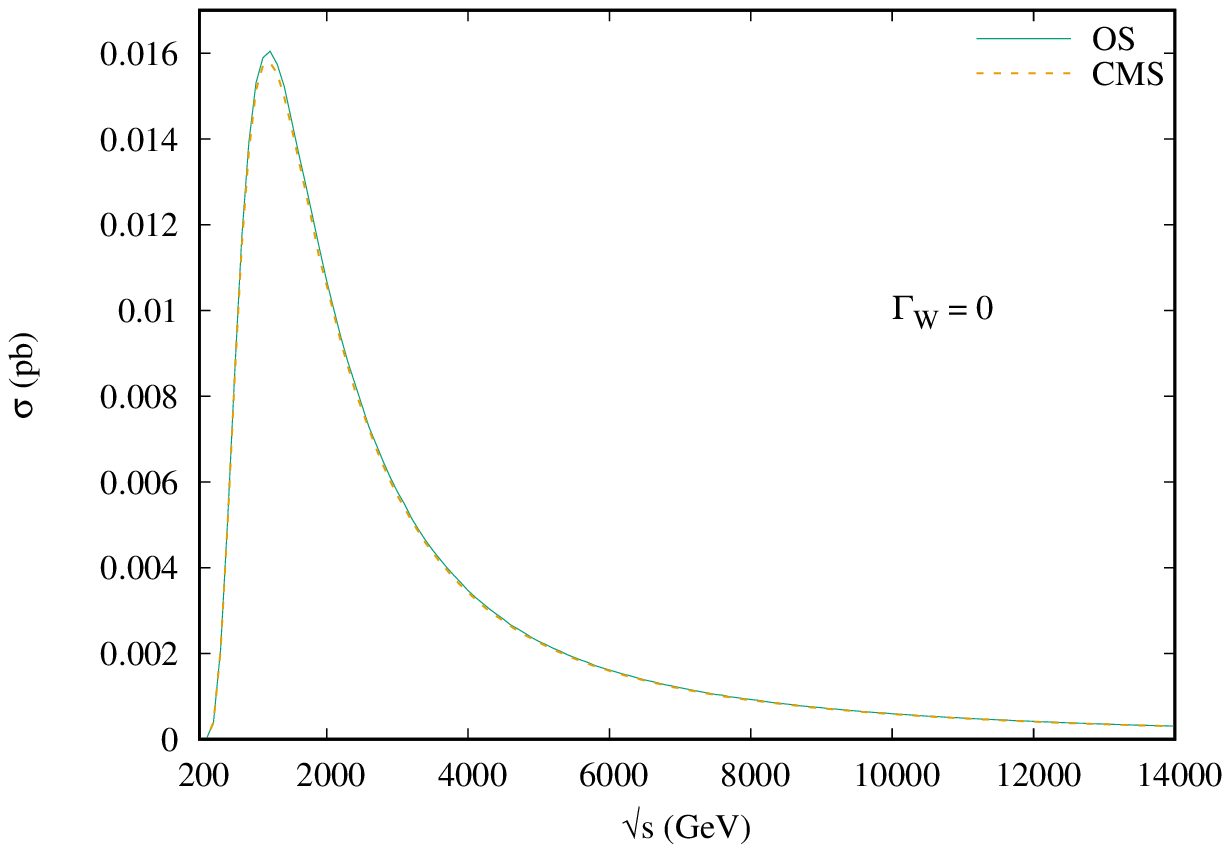}
	\includegraphics[scale=0.59]{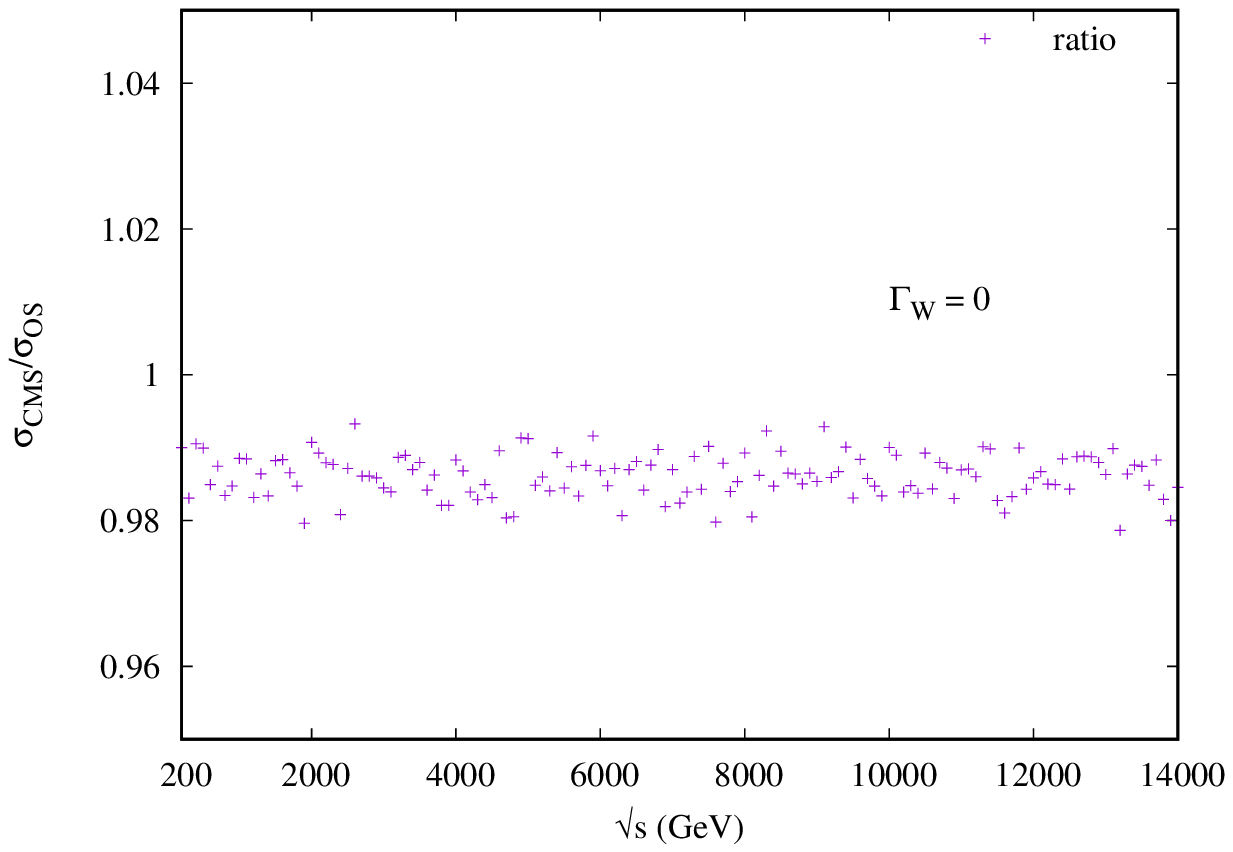}
\caption{Cross sections in OS and CMS at tree-level for $p p \to W_U W_U$ as a function of $\sqrt{s}$ for the case $\G_W = 0$ (left) and their ratio (right).} \label{fig:Born-GW0}
\end{figure}

\begin{figure}[h!]
	\centering
	\includegraphics[scale=0.59]{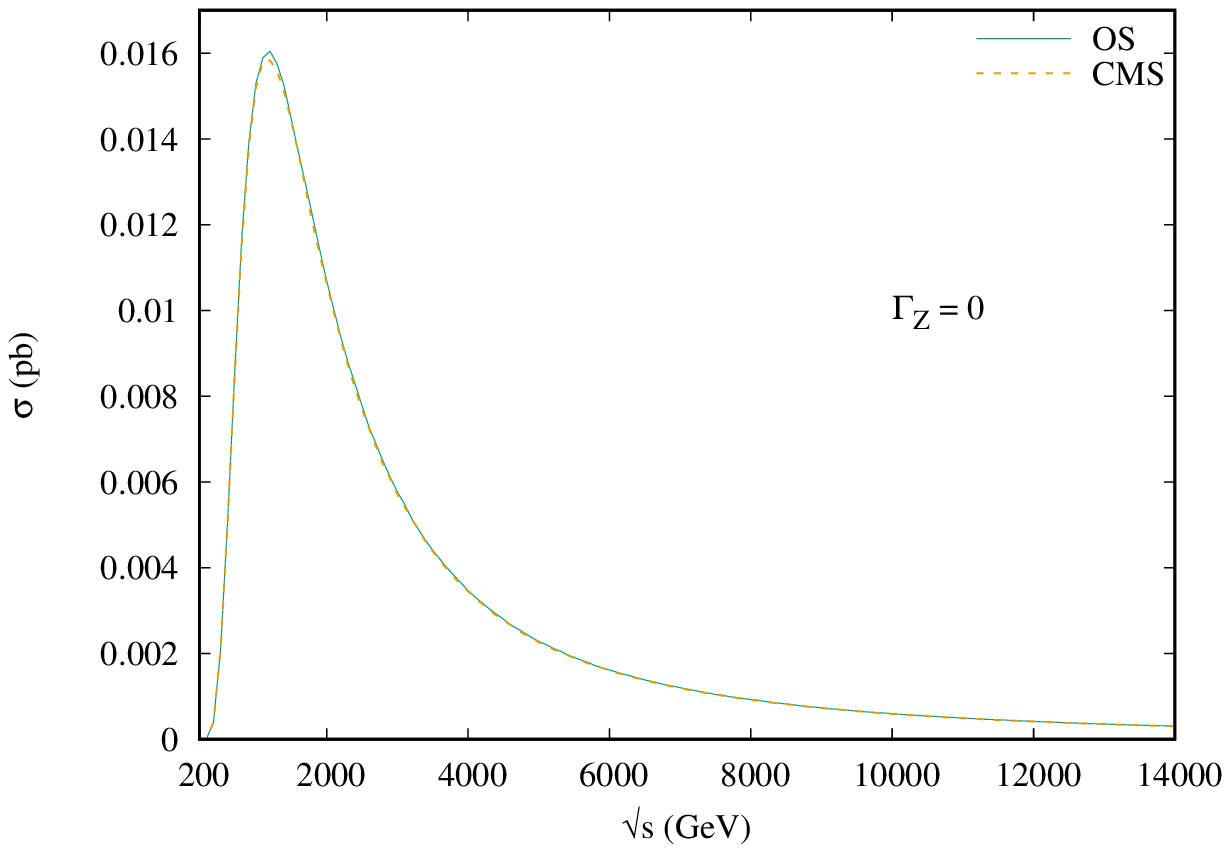}
	\includegraphics[scale=0.59]{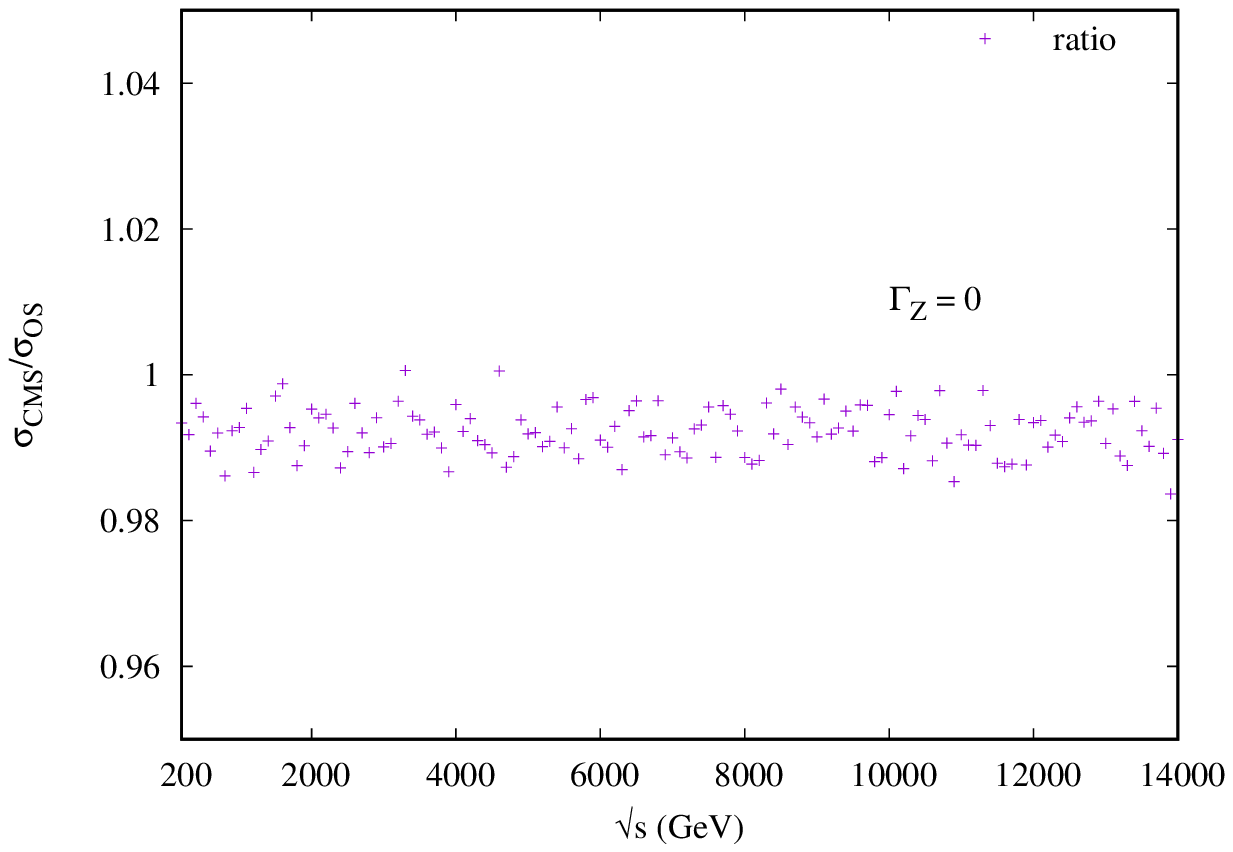}
\caption{Cross sections in OS and CMS at tree-level for $p p \to W_U W_U$ as a function of $\sqrt{s}$ for the case $\G_Z = 0$ (left) and their ratio (right).} \label{fig:Born-GZ0}
\end{figure}

In figures \ref{fig:Born-GH0} and \ref{fig:Born-GT0}, we see that relative to Fig. \ref{fig:Born-all} the widths $\G_H$ and $\G_t$ have almost no effect. This is due to the weak ratio $\G_H/M_H \sim 6 \times 10^{-3} \%$ and $\G_t/m_t \sim 0.7 \%$, although the latter is relatively higher. 
Fig. \ref{fig:Born-GW0} shows that the introduction of $\G_Z$ affects, both at low and high energies, the Born cross section at around $2 \%$. This effect has its origin in the relatively high value of the ratio $\G_Z/M_Z \sim 2.7 \%$. 
Moreover,  if we take into consideration the gauge boson $W$ as an unstable particle with a ratio $\G_W/M_w \sim 2.6 \%$, Fig. \ref{fig:Born-GZ0} shows that $\G_W$ has a comparable effect to that of $\G_Z$ if taken separately but opposite if combined together (see Fig. \ref{fig:Born-all}). 
In the next section we consider the CMS effect on the one-loop correction of the $p p \to  W^+ W^-$ process. 

\subsection{One-loop corrections (NLO)}

\begin{figure}[h!]
\centering 
\includegraphics[scale=0.8]{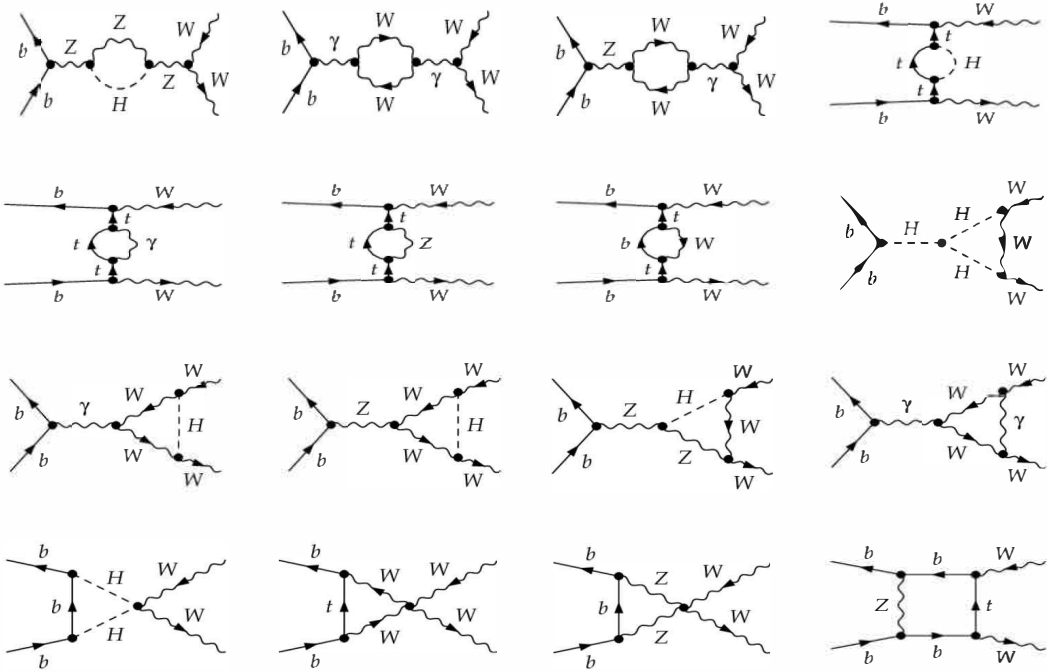}
\caption{A sample of one loop self-energy, vertex and box Feynman diagrams contributing to the process $p p \to WW$.} \label{fig:LoopFeynDiags}
\end{figure}

In Fig. \ref{fig:LoopFeynDiags}, we show a selected sample of self-energy, vertex and box diagrams, which are most sensitive to the widths of the unstable particles, among about 1000 diagrams contributing to the $pp \to WW$ process that we have generated using the fixed-order Monte Carlo program \texttt{MadGraph5\_aMC@NLO}. 
In the following, we present the results of NLO cross sections (Born + corrections) in both CMS and OS schemes for different widths of unstable particles. We consider in what follows below only the effect of the internal particle widths $\G_H$, $\G_Z$ and $\G_t$ on the one-loop corrections in the case where the external bosons are unpolarized and stable. 
This is for two reasons; one is linked to the fact that the Monte Carlo program \texttt{MadGraph5\_aMC@NLO} has not yet implemented one-loop polarizations for this type of process; the other reason is linked to Denner's remark that the external particles must be considered as stable in accordance with the work of Veltman \cite{VELTMAN1963186, Bouamerane-90}. We  therefore take the width of the $W$ boson, $\G_W$, to be  zero everywhere in the one-loop Feynman diagrams, since the same particle cannot be treated by two different renormalisation schemes simultaneously for the same process. 

It is worthwhile mentioning that our study does not involve the one-loop non-abelian QCD corrections. We only deal with the one-loop renormalisation of the electroweak contributions in both CMS and the usual OS schemes.  

\subsubsection*{Input parameters:}

For the purpose of our numerical investigations, we have used the following input values, in addition to those used for the tree level case \eqref{eq:Parameters}:
\begin{align}\label{eq:Parameters2}
m_e &= 0.51099906 \, \MeV,   \qquad  m_u = 47.0 \, \MeV,    \qquad  m_d = 47.0 \, \MeV, \notag\\
m_\mu &= 105.658389 \, \MeV,    \qquad  m_c = 1.55 \, \GeV,   \qquad     m_s = 150 \, \MeV, \notag\\
m_\tau &= 177.1 \, \MeV,   \qquad \;\;  \qquad  m_b = 4.5 \, \GeV.                               
\end{align} 
For the sake of deriving our results we have replaced the one-loop squared amplitude by the following formula \cite{Denner-1998}:   
\begin{equation}
|\mathcal{M}|^2 \to |\mathcal{M}_{Born}|^2 (1 + \delta_{soft}) +  2\, \mathcal{R}e\left(\mathcal{M}^*_{Born}\delta  \mathcal{M}\right), 
\end{equation}  
 where $\delta_{soft}$ takes into consideration soft bremstrahlung and is required in dealing with infrared divergences.  Moreover,  $\delta \mathcal{M}$ contains all one-loop Feynman diagrams as well as their corresponding counter-terms.

\subsubsection*{Results and discussion:}

\begin{figure}[h!]
	\centering
	\includegraphics[scale=0.59]{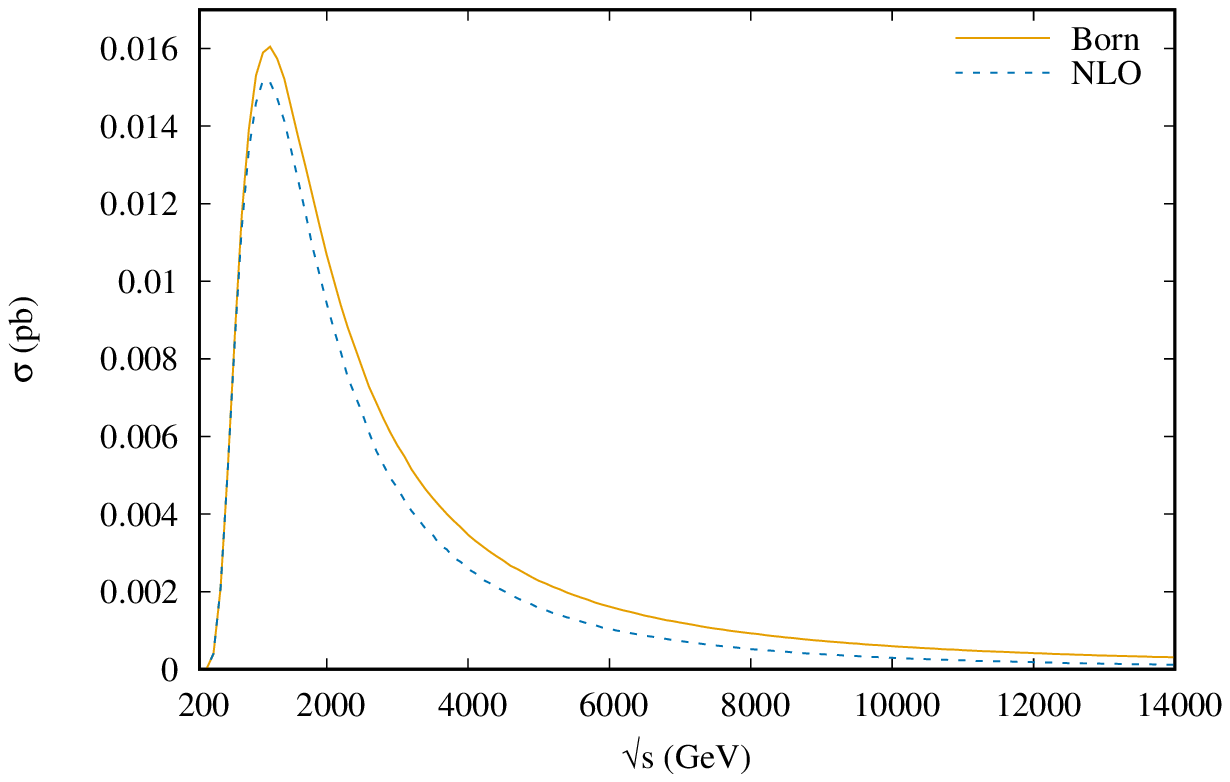}
	\includegraphics[scale=0.59]{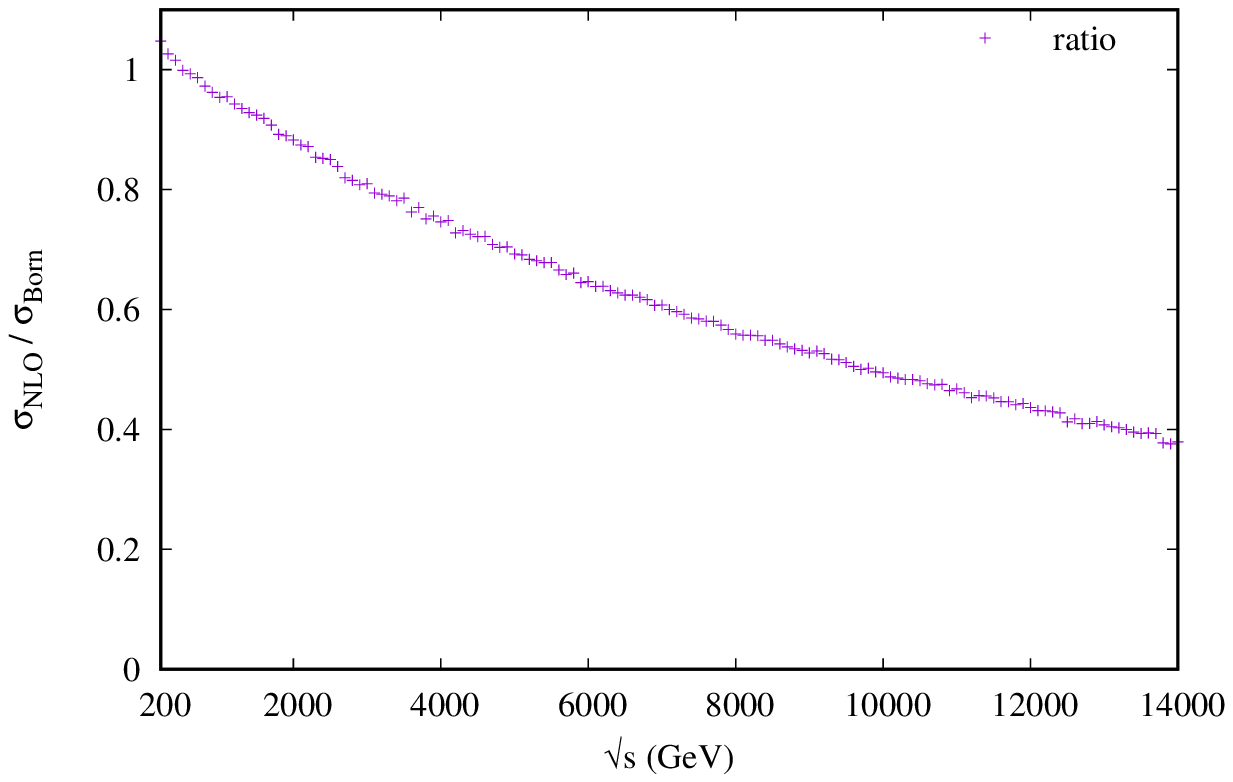}
	\caption{Plots for OS cross sections as a function of the scattering energy $\sqrt{s}$, comparing Born and NLO corrections.} \label{fig:NLO_SM}
\end{figure}

\begin{figure}[h!]
	\centering
v	\includegraphics[scale=0.57]{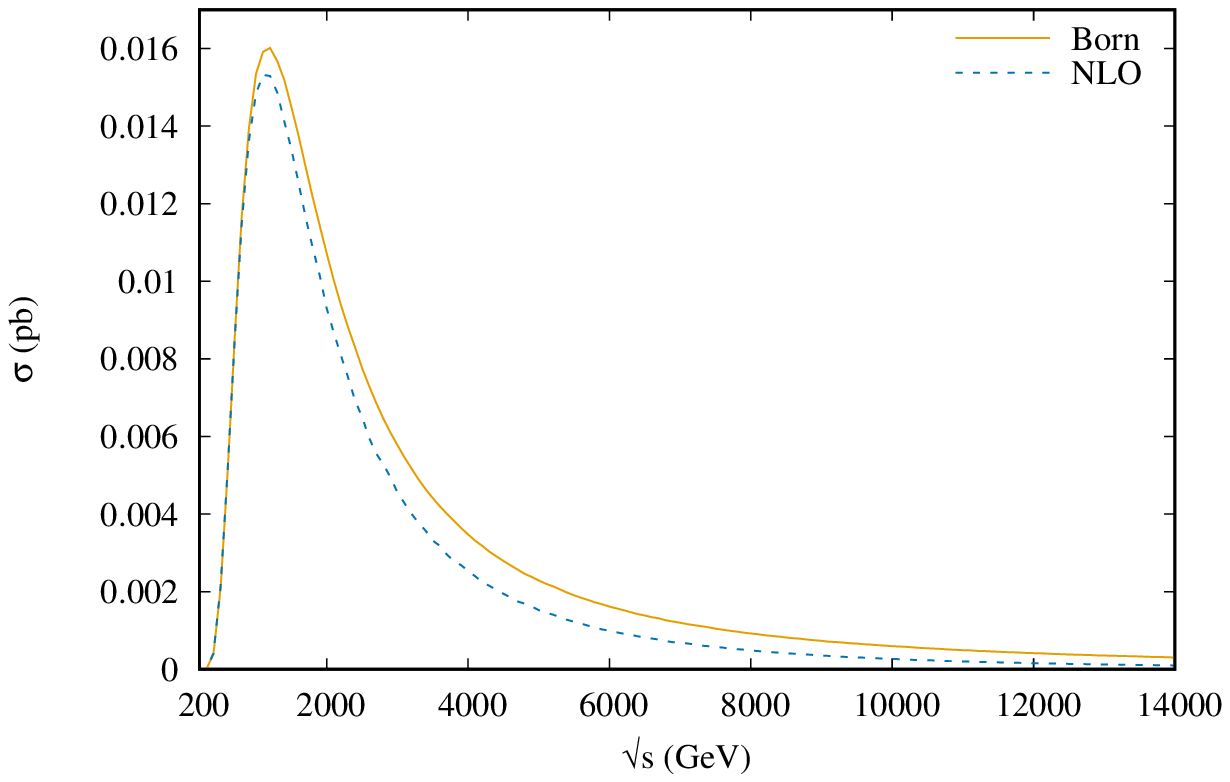}
	\includegraphics[scale=0.57]{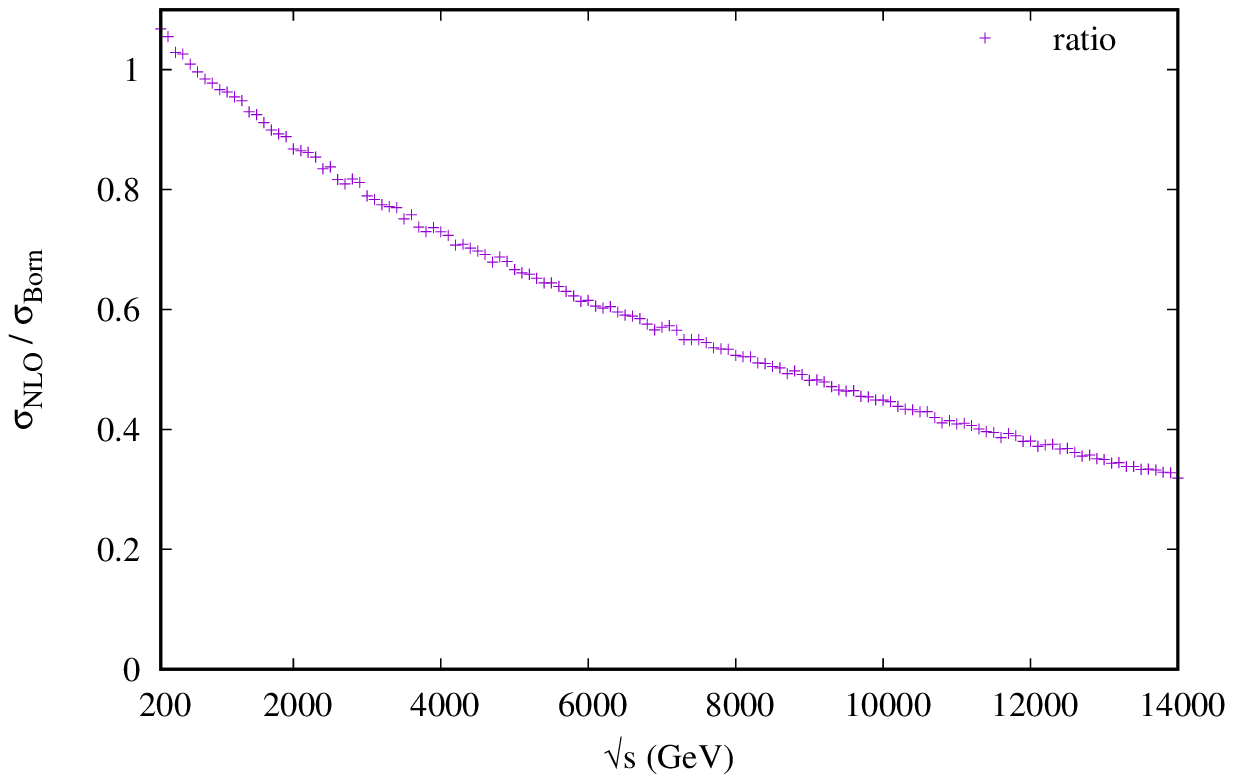}
	\caption{Plots for CMS cross sections as a function of the scattering energy $\sqrt{s}$, comparing Born and NLO corrections.} \label{fig:NLO_SMc}
\end{figure}

In figures \ref{fig:NLO_SM} and \ref{fig:NLO_SMc}, we present, for stable and unpolarized external bosons, the full  cross sections at one-loop in OS and CMS schemes respectively. At high energies, they behave like $1/s$, which confirms that the two schemes do not break the unitarity of the one-loop results. At low energies, the corrections are positive and of the order of $5 - 10 \,\%$ in both schemes. Once the full NLO cross sections reach their maxima, around $1000\, \GeV$, the corrections become negative and large. They maintain this behaviour throughout the remaining energy range. In the range $2000 - 6000\, \GeV$ the corrections are around   $10 - 30\,\%$ in OS and $10 - 40_,\%$ in CMS. Furthermore, they reach about $55\,\%$ in OS and $65\,\%$ in CMS around $14\,$TeV.
Thus at high energies, we notice that the corrections are of the order of the Born cross section in both schemes. Hence the $\mathcal{O}(\alpha^2)$  corrections should be taken into consideration in the calculation of the corrected cross section, in order to obtain more reliable and precise results.  

\begin{figure}[h!]
	\centering
	\includegraphics[scale=0.59]{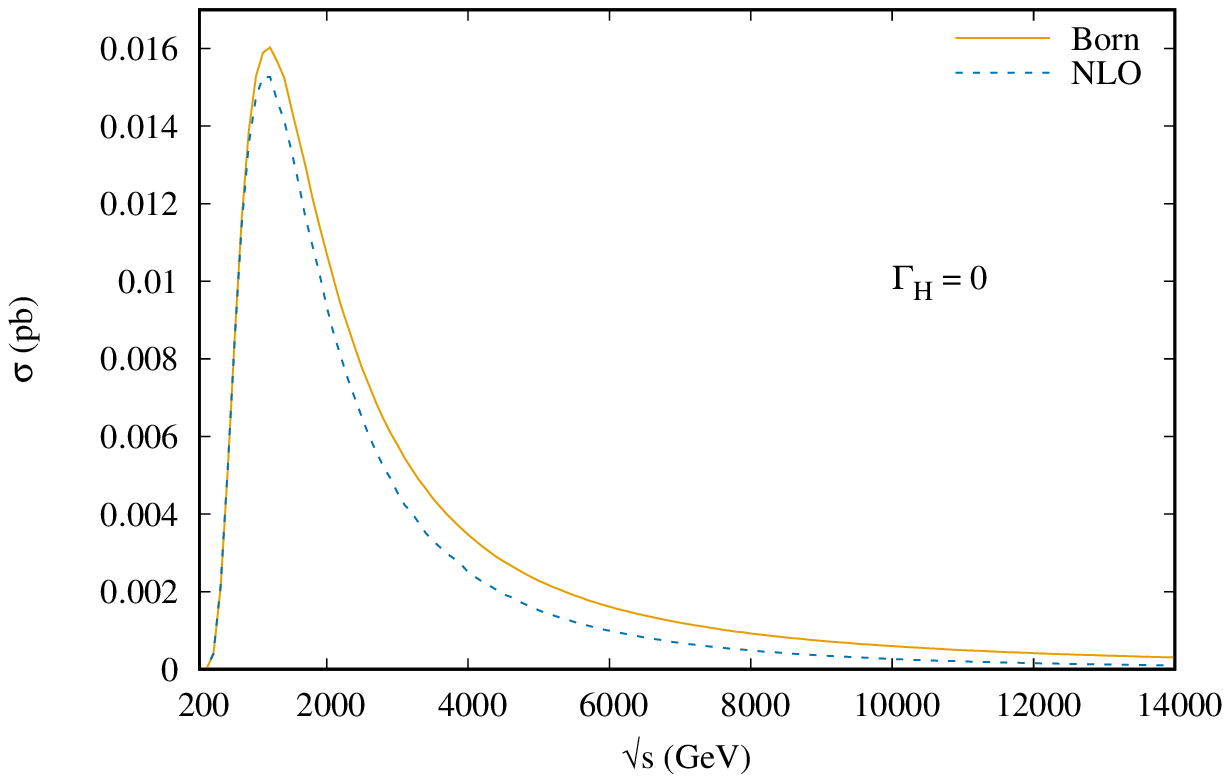}
	\includegraphics[scale=0.59]{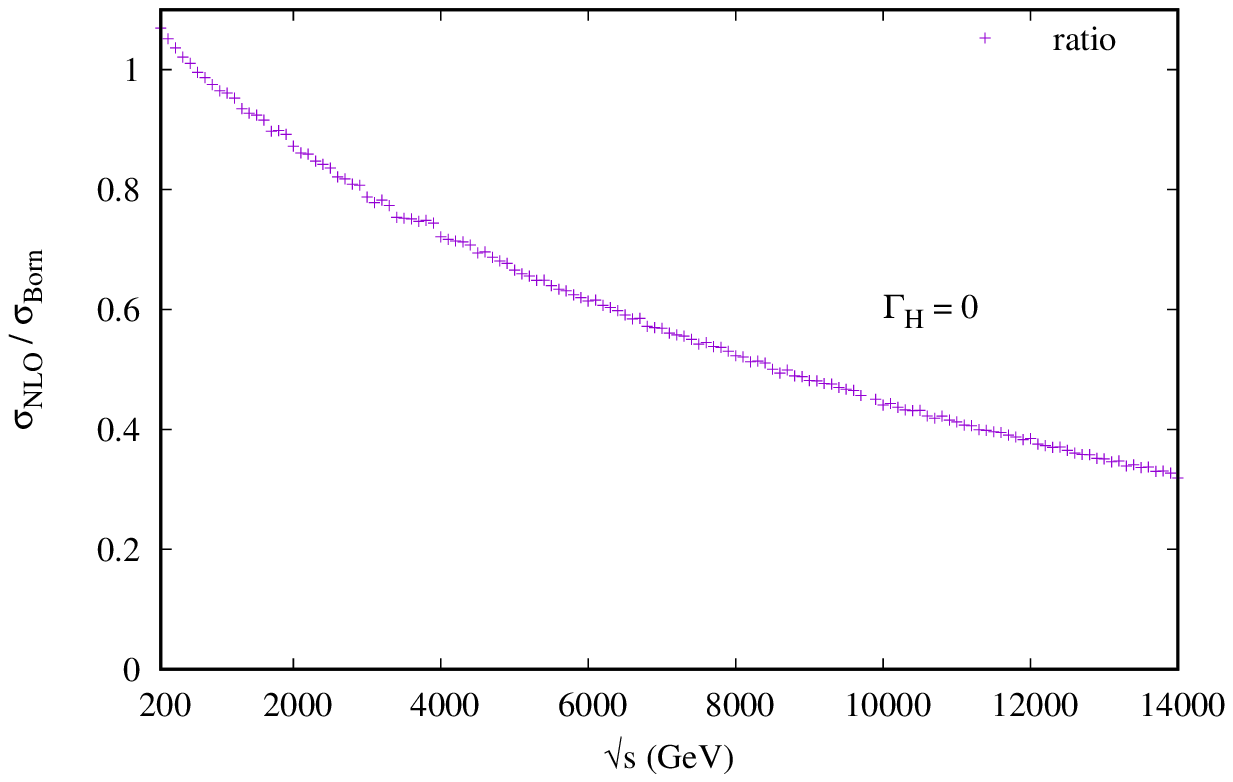}		
	\caption{Plots for CMS cross sections as a function of the scattering energy $\sqrt{s}$, comparing Born and NLO corrections, for the case $\G_H = 0$.} \label{fig:NLO_SM-GH0}
\end{figure}

\begin{figure}[h!]
	\centering
	\includegraphics[scale=0.59]{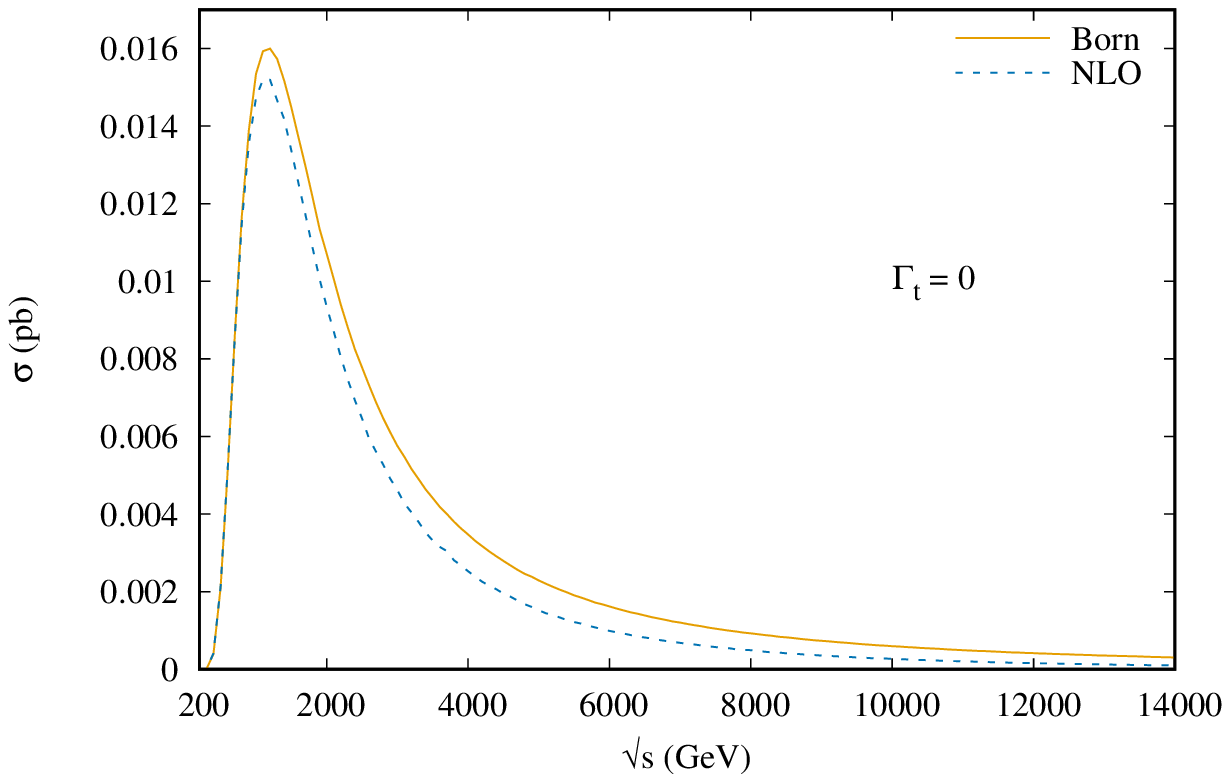}
	\includegraphics[scale=0.59]{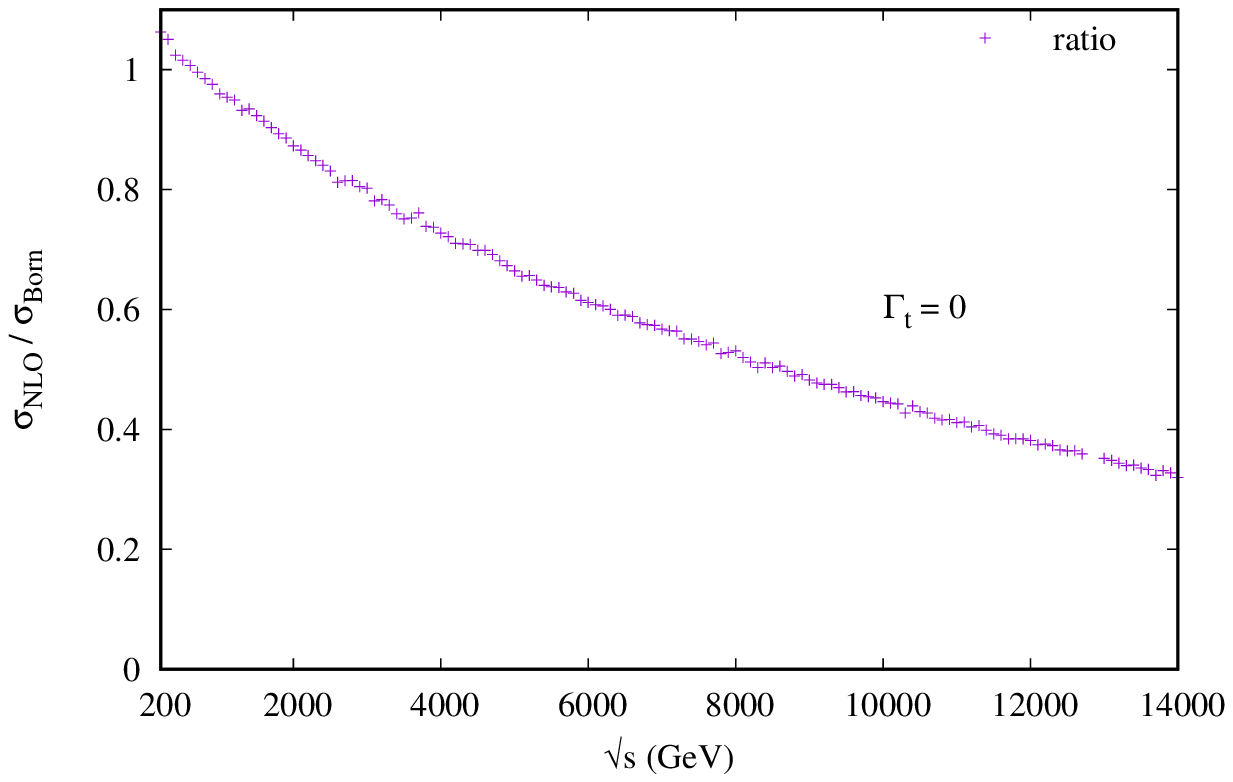}		
	\caption{Plots for CMS cross sections as a function of the scattering energy $\sqrt{s}$, comparing Born and NLO corrections, for the case $\G_t = 0$.} \label{fig:NLO_SMc-GT0}
\end{figure}

Figures \ref{fig:NLO_SM-GH0} and \ref{fig:NLO_SMc-GT0} show that only $\G_Z $ significantly affects the corrections at one-loop.  Whereas $\G_H$ and $\G_t$ present insignificant effects, even if the former preserves its ability to restore unitarity as in the case of the actual OS scheme. 

\begin{figure}[h!]
	\centering
	\includegraphics[scale=0.59]{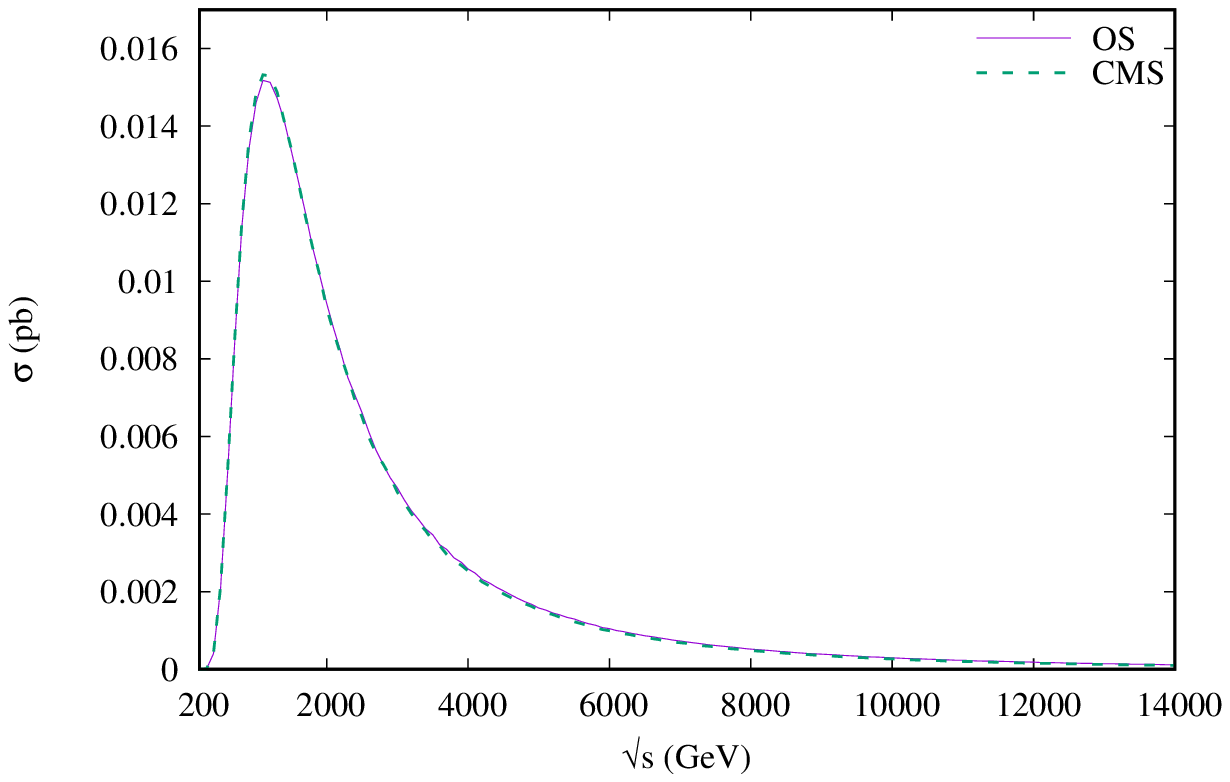}
	\includegraphics[scale=0.59]{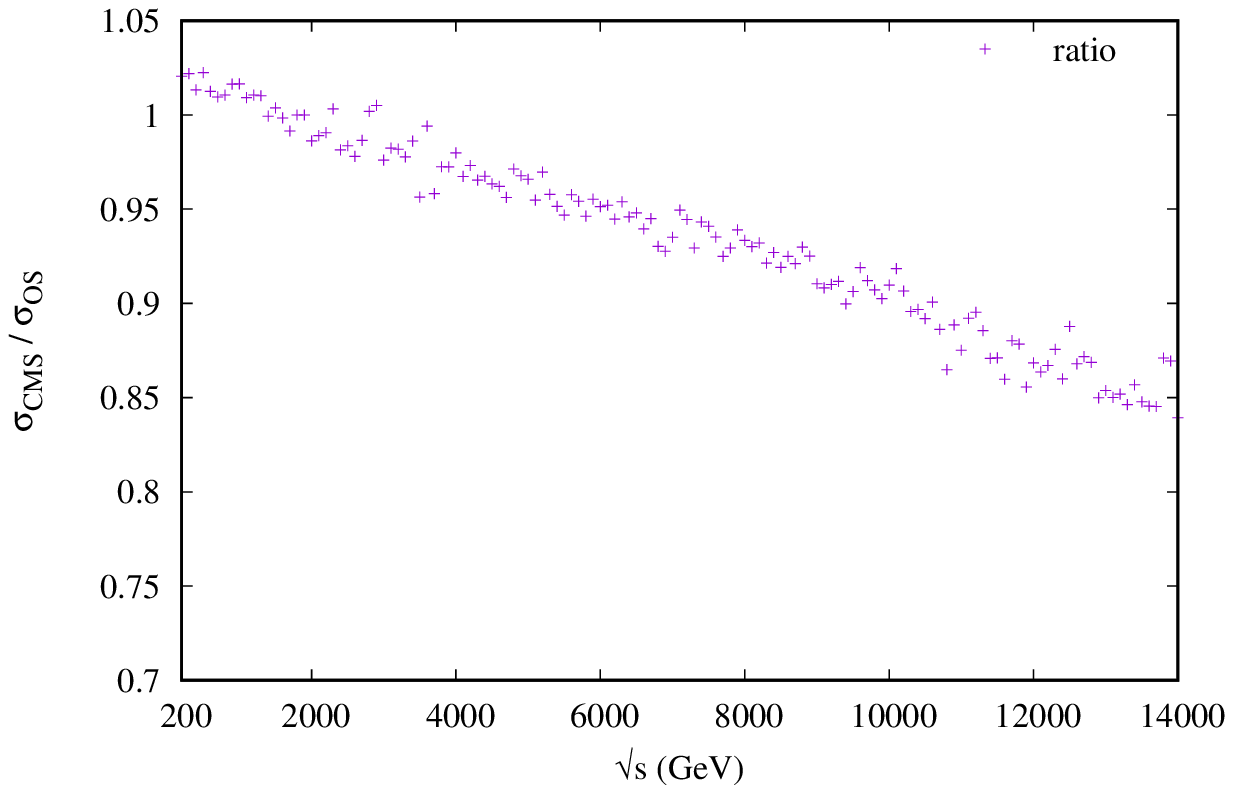}
	\caption{Plots for SM (real) and SMc (CMS) cross sections as a function of the scattering energy $\sqrt{s}$ at NLO.} \label{fig:CMS-SM-all}
\end{figure}

Figure \ref{fig:CMS-SM-all} shows a comparison between the full cross sections in the two aforementioned schemes.  At low energy, the effect of the width $\G_Z$  on  the full cross section is around $2.5\,\%$, which vanishes around $1000\, \GeV$ and then increases with energy until it reaches around $15\,\%$ at $14\,$ TeV. This behaviour is expected to hold for further high energies. That is,  the effect of $\G_Z$ increases with energy. This, however,  remains to be verified, especially if this process is considered as an internal part of another process with stable external particles. This is to be investigated in our future work.  

\section{Conclusion}
\label{sec:Conclusion}

In the present work, we made an extension of the calculations of the electroweak radiative corrections to a loop of the process $ pp\to WW$ to the complex mass scheme, where the widths of the unstable particles are introduced respecting the identity of Ward and that of Slavnov- Taylor.

At tree level, we obtained for the longitudinal mode of the external gauge bosons a total amplitude proportional to the width $\G_Z$ of the internal gauge boson Z. We have shown that  upon the inclusion of the width $\G_W$ of the external $W$ gauge boson, the total amplitude depends on both $\G_W$ and $\G_Z$. The other widths $\G_H$ and $\G_t$ of the Higgs boson and the top quark have practically no effect on neither the amplitude nor the Born cross section. The effect of $\G_Z$ on the Born cross section is about $2\,\%$ for unpolarized and stable W bosons. When the width $\G_W$ of the $W$ external gauge boson is included, it produces an effect on the Born cross section comparable to that of $\G_Z$ if they are introduced separately, and an opposite effect to $\G_Z$ if they are introduced together.

The Born and full one-loop cross sections behave as $1/s$ at high energies in both CMS and OS schemes, thus preserving the unitarity of the theory. The one-loop corrections are affected by $\G_Z$ up to $65\%$ in the CMS scheme at energy of about $14\,$ TeV. They reach for the same case about $55\%$ in the real OS scheme. They are therefore of the order of the Born cross section (at high energy), and  hence higher orders, $\mathcal{O}(\alpha^2)$ and above,  have to be included for any meaningful predictions.

Finally, comparisons of the full cross sections in OS and CMS schemes, revealed that the width $\G_Z$ affects the NLO cross section by about $5\%$ around 2 TeV and $15\%$ around 14 TeV. This effect therefore increases with increasing energy and its behaviour after 14 TeV remains to be verified. For a full study of this type of problems where the effect  of the width $\G_W$ is not neglected, as well as the widths of the other unstable  particles, $\G_Z, \G_H$ and $\G_t$, on the one-loop corrections, it suffices to relate this process, i.e, $ p p \to WW$, to another one where the external particles are stable.  Another important point to consider for future work is the effect of various widths in the case where polarisation is not neglected in tree and one-loop calculations.

\acknowledgments

We would like to thank Dr. Noureddine Bouayed for support and encouragement to carry out this work, valuable discussions particularly on complex renormalisation as well as collaboration on the usage of \texttt{FeynCalc}.

\appendix

\section*{Appendix}

\section{Amplitudes in CMS}
\label{app:Amps_CMS}

The different high-energy amplitudes at the tree level in CMS in the case where the longitudinal external gauge bosons are stable and unstable are given by the following expressions:
\begin{align}\label{eq:Amps_HEL_CMS1}
\Re\{\cM_\g^s\} &= -Q_f \, e^2 \, \frac{\vb(p_2)\,\ \slashed{q_1} u(p_1)}{M_W^2}+ \cO(1/x),\qquad\qquad\qquad\qquad\qquad\qquad\qquad\qquad\qquad\quad\, \notag \\
\Im\{\cM_\g^s\} &= 0 + \cO(1/x). 
\end{align}
\vspace*{-0.75cm}
\begin{align}\label{eq:Amps_HEL_CMS2}
\Re\{\cM_Z^s\} &= e^2 \, \frac{\vb(p_2)}{M_W^2} \Big[ Q_f \,\ \slashed{q_1}+ \frac{M_Z^2 \,\ T_3^f}{2 (M_Z^2 - M_W^2)} \Big( m_q (1 - 2 P_L) - 2 \slashed{q_1} P_L \Big) \Big] u(p_1) + \cO(1/x), \quad\, 
\notag\\
\Im\{\cM_Z^s\} &= e^2 \, \frac{\G_Z \,\ M_Z \,\ T_3^f}{2(M_Z^2- M_W^2)^2} \,\ \vb(p_2) \Big( m_q (1-2P_L)-2 \slashed{q_1} P_L \Big) u(p_1)+ \cO(1/x).
\end{align}
\vspace*{-0.5cm}
\begin{align}\label{eq:Amps_HEL_CMS3}
\Re\{\cM_q^s\} &= e^2 \, \frac{- M_Z^2}{2 M_W^2 (M_Z^2- M_W^2)} \,\ \vb(p_2) \Big( m_q + \slashed{q_1} \Big) P_L u(p_1)+ \cO(1/x),\qquad\qquad\qquad\qquad\quad\, \notag\\
\Im\{\cM_q^s\} &= e^2 \, \frac{- \G_Z M_Z}{2 (M_Z^2- M_W^2)^2} \,\ \vb(p_2) \Big( m_q + \slashed{q_1} \Big) P_L u(p_1)+ \cO(1/x).
\end{align}
\vspace*{-0.5cm}
\begin{align}\label{eq:Amps_HEL_CMS4}
\Re\{\cM_H^s\} &= e^2 \, \frac{m_q  M_Z^2}{4 M_W^2 (M_Z^2- M_W^2)} \,\ \vb(p_2) u(p_1)+ \cO(1/x),\qquad\qquad\qquad\qquad\qquad\qquad\quad\quad\,\, \notag\\
\Im\{\cM_H^s\} &= e^2 \, \frac{ \G_Z m_q  M_Z}{4 (M_Z^2- M_W^2)^2} \,\ \vb(p_2) u(p_1)+ \cO(1/x).
\end{align}
\begin{align}\label{eq:Amps_HEL_CMS5}
\Re\{\cM_\g^s\} &= -Q_f \, e^2 \,  \frac{\vb(p_2)\,\ \slashed{q_1} u(p_1)}{M_W^2}+ \cO(1/x),\qquad\qquad\qquad\qquad\qquad\qquad\qquad\qquad\qquad\quad\, 
\notag \\
\Im\{\cM_\g^s\} &= -Q_f \, e^2 \, \frac{\G_W}{M_W} \, \frac{\vb(p_2) \slashed{q_1} u(p_1)}{M_W^2} + \cO(1/x).
\end{align}
\vspace*{-0.7cm}
\begin{align}\label{eq:Amps_HEL_CMS6}
\Re\{\cM_Z^s\} &= e^2 \,  \frac{\vb(p_2)}{M_W^2} \Big[ Q_f \,\ \slashed{q_1}+ \frac{M_Z^2 \,\ T_3^f}{2 (M_Z^2 - M_W^2)} \Big( m_q (1 - 2 P_L) - 2 \slashed{q_1} P_L \Big) \Big] u(p_1) + \cO(1/x),\,\,\,\,\, \notag\\
\Im\{\cM_Z^s\} &= e^2 \, \frac{M_Z \, T_3^f}{2( M_Z^2-M_W^2)^2} \, \vb(p_2) \Biggl\{ \G_Z + \frac{\G_W}{M_W} \, \frac{M_Z \left( M_Z^2-2 M_W^2 \right)}{M_W^2} \Biggr\} \Biggl\{ m_q \left(1-2P_L \right)
\notag\\
&   - 2 \slashed{q_1} P_L \Biggr\} \, u(p_1) + Q_f \frac{\G_W}{M_W} \, \frac{\left(M_Z^2- 2 M_W^2\right)}{M_Z^2-M_W^2} \, \slashed{q_1} \,\vb(p_2) \, u(p_1) + \cO(1/x).
\end{align}
\vspace*{-0.5cm}
\begin{align}\label{eq:Amps_HEL_CMS7}
\Re\{\cM_q^s\} &= e^2 \, \frac{- M_Z^2}{2 M_W^2 (M_Z^2- M_W^2)} \,\ \vb(p_2) \Big( m_q + \slashed{q_1} \Big) P_L u(p_1)+ \cO(1/x),\qquad\qquad\qquad\qquad\,\, \notag\\
\Im\{\cM_q^s\} &= e^2 \,  \frac{-M_Z}{2 \left(M_Z^2-M_W^2 \right))^2} \, \vb(p_2) \Biggl\{ \G_Z + \frac{\G_W}{M_W} \, \frac{M_Z \left( M_Z^2- 2 M_W^2 \right)}{M_W^2} \Biggr\}
\notag\\
& + \left(m_q + \slashed{q_1} \right) P_L u(p_1) + \cO(1/x).
\end{align}
\vspace*{-0.8cm}
\begin{align}\label{eq:Amps_HEL_CMS8}
\Re\{\cM_H^s\} &= e^2 \, \frac{m_q  M_Z^2}{4 M_W^2 (M_Z^2- M_W^2)} \,\ \vb(p_2) \, u(p_1)+ \cO(1/x),\qquad\qquad\qquad\qquad\qquad\qquad\qquad
 \notag\\
\Im\{\cM_H^s\} & = e^2 \, \frac{\G_Z \, m_q \, M_Z}{4 \left(M_Z^2-M_W^2 \right)^2} \, \vb(p_2) \, u(p_1) + \cO(1/x).
\end{align}
\newpage
\bibliographystyle{JHEP}
\bibliography{CMS-refs}

\end{document}